\documentclass[onecolumn,journal]{IEEEtran}
\makeatletter

\def\normalsize{\@setfontsize{\normalsize}{10}{11.4pt}}%
\normalsize

\makeatother

\usepackage{bm}
\usepackage{url}
\usepackage{tabularx}
\usepackage{booktabs}
\usepackage{graphicx}
\usepackage[cmex10]{amsmath}
\usepackage[font=footnotesize]{subcaption}
\usepackage[font=footnotesize,justification=centering]{caption}
\usepackage[noadjust]{cite}
\usepackage{float}
\usepackage{color}
\usepackage{soul} %to strike out text
\usepackage{flushend}
\usepackage{amsfonts}
\usepackage[colorlinks,linkcolor=black,anchorcolor=black,citecolor=black]{hyperref}
\DeclareCaptionLabelSeparator{periodspace}{.\quad}
\graphicspath{{Figures/}}

\usepackage{amsfonts}
\usepackage{amssymb}
\newtheorem{assumption}{Assumption}

\newtheorem{proposition}{Proposition}

\newtheorem{remark}{Remark}

\def\L2e{{\cal L}_{2e}}

\def\begequarrs{\begin{eqnarray*}}
	\def\endequarrs{\end{eqnarray*}}
\def\begequarr{\begin{eqnarray}}
\def\endequarr{\end{eqnarray}}
\def\begarr{\begin{array}}
	\def\endarr{\end{array}}
\def\begequ{\begin{equation}}
\def\endequ{\end{equation}}

\def\begdes{\begin{description}}
	\def\enddes{\end{description}}
\def\begenu{\begin{enumerate}}
	\def\begite{\begin{itemize}}
		\def\endite{\end{itemize}}
	\def\endenu{\end{enumerate}}

\def\lef[{\left[\begin{array}}
	\def\rig]{\end{array}\right]}

\def\begcen{\begin{center}}
	\def\endcen{\end{center}}
\def\begrem{\begin{remark}\rm}
	\def\endrem{\end{remark}}
\def\begcas{\begin{cases}}
	\def\endcas{\end{cases}}%
\ifCLASSINFOpdf
\else
\fi
\begin{document}
\title{Updated version ``Robust Voltage Regulation of DC-DC Buck Converter With ZIP Load via An Energy Shaping Control Approach"}

\author{
	\vskip 1em
	{
	Wei He,
    Yanqin Zhang,
	Yukai Shang,
    Mohammad Masoud Namazi,
     Wangping Zhou,
     Josep M. Guerrero
	}

	\thanks{
		
		{
		Wei He, Yanqin Zhang, Yukai Shang and Wangping Zhou are with Nanjing University
of Information Science and Technology, Nanjing 210044, China.

       Josep M. Guerrero is with the Center for Research on Microgrids (CROM), AAU Energy, Aalborg University, 9220 Aalborg East, Denmark.
		}
	}
}
\maketitle

%\begin{abstract}
%For multi-converter power electronic systems, the load converters are always considered as the constant power loads (CPLs) of the feeder converters. It is known that the dynamic behavior of CPLs exhibits negative impedance characteristic, which may lead to the instability of overall power systems. In this paper, the output voltage regulation problem of DC-DC power converters subject to unknown CPL is investigated by using adaptive energy shaping control (ESC) algorithm. A new immersion and invariance observer is proposed to estimate the extracted power load. It is ensured that the closed-loop system is (locally) asymptotically stable with a guaranteed domain of attraction. Finally, the detail simulation and experimental results are carried out to validate the performance of the proposed methods.
%\end{abstract}
%
%% Note that keywords are not normally used for peerreview papers.
%\begin{IEEEkeywords}
%DC-DC power converters, constant power load (CPL), energy shaping control (ESC), interconnection and damping assignment passivity based control (IDA-PBC), immersion and invariance (I\&I).
%\end{IEEEkeywords}

% \fi
\IEEEpeerreviewmaketitle
\begin{abstract}
ZIP loads (the parallel combination of constant impedance loads, constant current loads and constant power loads) exist widely in power system. In order to stabilize buck converter based DC distributed system with ZIP load, an adaptive energy shaping controller (AESC) is devised in this paper. Firstly, based on the assumption that lumped disturbances are known, a full information controller is designed in the framework of the port Hamiltonian system via energy shaping technique. Besides, using mathematical deductive method, an estimation of the domain of attraction is given to ensure the strict stability. Furthermore, to eliminate the influence of parameter perturbations on the system, a disturbance observer is proposed to reconstruct the lumped disturbances and then the estimated terms are introduced to above controller to form an AESC scheme. In addition, the stability analysis of the closed-loop system is given. Lastly, the simulation and experiment results are presented for assessing the designed controller.
\end{abstract}
\begin{IEEEkeywords}
DC-DC buck converter, ZIP load, energy shaping, disturbance observer, adaptive control
\end{IEEEkeywords}
\section{Introduction}
\subsection{Literature review}
\begin{figure}
	\centering
	\includegraphics[scale=0.45]{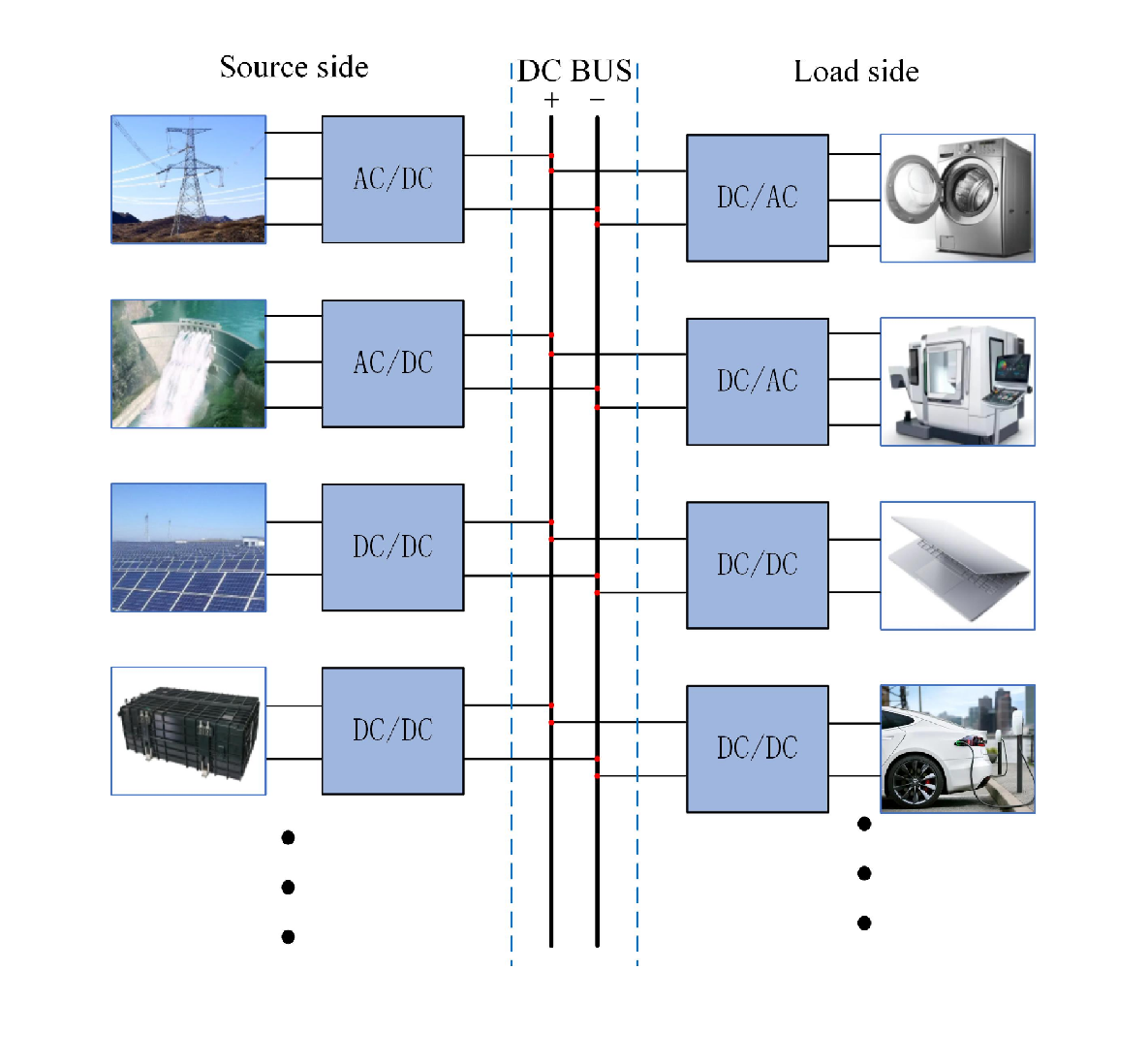}
\vspace{-0.5cm}
	\caption{A typical structure of DC microgrids.}
	\label{Fig:DCMG}
\end{figure}

DC-DC converters are often adopted as an interface between power supply and load to provide an appropriate working condition in practical applications, such as microgrid, ship power system, transportation vehicles, microprocessors \cite{hossain2018recent,nizami2016intelligent,he2019design}. {A typical structure of microgrid system containing DC-DC converters is shown in Fig. \ref{Fig:DCMG}} \cite{hassan2018adaptive}. {It is observed that new energy resources are regarded as generation unites, which provide power for  domestic electrical installation and other applications. Wherein, DC-DC converters are installed to regulate the voltage of the ports between generation units and DC bus as well as DC bus and electrical equipments, which play an important role in power conversion.} Hence, their control performance definitely affect the power quality of overall system \cite{zhou2013constant}. It is noted that buck converter, which is one of an important part in the converters, is adopted to step down the voltage of output port \cite{chen2021current}. It is necessary that an effective control strategy should be designed to regulate its output voltage with nice transient and steady state performance.

For the case of classical resistance load, its average model can be constructed as a second-order linear system. Obviously, there are a lot of remarkable results for dealing with the regulation problem of output voltage of buck converter under this kind of load in literature. For example, a switching control scheme based on energy conservation principle is reported in \cite{7405331} to deal with the regulation problem of output voltage of the system. In \cite{komurcugil2012adaptive}, the authors propose an adaptive terminal sliding-mode control for the system.  Besides, a second-order sliding-mode control scheme is developed with the consideration of model uncertainties and external disturbances \cite{7104150}.  It is developed in \cite{8948314} that an event-triggered model predictive controller is designed for it, which captures less computational burden and less switching actions. A simple current constrained control approach is proposed for the system to ensure that the inductor current does not go beyond the limit \cite{9686600}. In this case, a higher reliability and safety can be obtained. In \cite{sira2010robust}, the authors propose a GPI observer-based robust control scheme to regulate output voltage around the equilibrium and reduce the affect of time-varying disturbances caused by load variation and input voltage fluctuation on the system, which obviously enhances the robustness. In \cite{hernandez2009adaptive}, an adaptive proportional-integral passivity-based controller (PI-PBC) without needing the information of the load is proposed for DC-DC converters on the basis of immersion and invariance technique. Additionally, a robust P leaky ID-PBC is developed for the system, which ensures that the system is exponentially stable and shows a high robustness against variable operating conditions \cite{zonetti2022pid}. To further improve the transient performance, an adaptive finite-time control method is developed for the system in the presence of unknown input voltage and load variations, which captures a faster regulation performance and stronger robustness \cite{cheng2017adaptive,zhang2020universal,chen2021current,nizami2018analysis}. The reviews about this topic can be found in \cite{hossain2018recent}. With the increasing demand for flexible regulation, cascaded DC-DC converters exist in power system \cite{he2021stabilization}. Internal interaction between the closely regulated converters will cause the appearance of constant power load (CPL).

For the case of CPL, some advanced control methods are proposed to compensate negative impendence caused by it and ensure the stability of the system. In \cite{he2019design,he2021voltage}, an adaptive energy shaping controller is devised for stabilizing buck converter with CPL, which obtains a satisfactory performance even if the load information is unavailable. In addition, to ensure the strict convergence, the estimation of attraction domain is presented. It is reported in \cite{hassan2018adaptive} that an adaptive passivity based controller is designed for the system by integrating with a nonlinear disturbance observer. A nonlinear controller based on feedback linearization technique is proposed in \cite{solsona2015nonlinear}, in which a nonlinear reduced order observer is designed for estimating the power load. Besides, an improved interconnection and damping assignment passivity based control scheme is proposed by developing an adaptive interconnection matrix \cite{8818313}. An offset-free model predictive controller is proposed for the system with a rigorous stability analysis by designing a higher order sliding mode observer, which shows a nice anti-disturbance performance under load variation and system uncertainties \cite{8839850}. In \cite{boukerdja2020h}, the authors present a H$_{\infty}$ based robust control scheme to deal with the control problem of the system. In \cite{9186703}, the authors develop a composite robust discretized quasi-sliding mode control scheme for stabilization of buck converter fed dc microgrids with CPLs on the basis of large signal stability, which improves disturbance rejection and inherent chattering suppression. The authors in \cite{mayo2020power} proposes a power shaping control scheme for regulation problem of output voltage of the system. Designing a reduced-order generalized parameter estimation-based observer, an adaptive sensorless control scheme is developed for the system \cite{he2022adaptive}. The implementation of the controller does not need the information of inductor current and power load. A deep learning based voltage controller is developed in \cite{gheisarnejad2020novel,gheisarnejad2020,hajihosseini2020dc}, which simples the control design. However, it results in high computational burden and requires high performance processors. The reviews about this topic can be found in \cite{singh2017constant,xu2020review}. Note that many loads in power systems, such as electronic devices and appliances, exhibit a nonlinear behavior. It is worth noting that ZIP load model allows for the representation of this non-linear behavior and accurately captures the load characteristics under different operating conditions, which is essential for ensuring accurate representation and analysis of the behavior and impact of various types of loads.

In the case, it is developed in \cite{bahrami2023large} that an adaptive nonlinear controller is proposed for buck converter with ZIP load by combining backstepping controller with adaptation law. Besides, a nonlinear controller is proposed for buck converter with ZIP load based distributed generation units \cite{bahrami2023decentralized}.  A sliding mode controller for buck-boost converter with ZIP load is developed in \cite{singh2016mitigation}. The influence of the perturbations of composite load and input voltage on the system may be reduced by integral action and sgn function.  In \cite{cucuzzella2019robust}, a robust passivity based control (RPBC), which doesn't depend on load information, is developed for boost converter with ZIP load. But the implementation of the controller needs first-time derivatives of current and voltage. In \cite{sadabadi2020distributed}, a distributed controller without the requirement of load information is designed for parallel buck converter with ZIP load. The simulation results show that it has a nice robustness against load variation and uncertainty of the capacitance.

\subsection{Contribution}
In this paper, using the framework of port-Hamiltonian (PH) system, an adaptive energy shaping controller (AESC) is designed for buck converter based a DC distribution system  with ZIP load. {The contributions are given as follows.

\begin{enumerate}
	\item[1)] By constructing a reference dynamic system with the consideration of time-varying disturbances, an AESC scheme is proposed by describing the system as a PH form.
	
	\item[2)] In our method, a disturbance observer is designed to estimate time-varyingly matched and mismatched disturbances consisting of parameter perturbations, unlike some existing control strategies, in which the exact information of some circuit parameters are required.  Satisfaction of this requirement may be hard in practice due to a dynamic or unpredictable environment.
	
	\item[3)] Using mathematical induction method, a well-defined domain of attraction is given to ensure the strict stability, unlike the existing results in the literature, in which a domain of attraction fails to be given to guide the selection of initial condition. A comparison study among the proposed controller, PI control and a passivity based control method is carried out via simulation and experimental results.
\end{enumerate}
}

The remaining parts are organized as follows. In Section \ref{sec2}, the state space model of buck converter with ZIP load based a DC distribution system is provided and the control objectives are described. Section \ref{sec3} presents the controller design. Section \ref{sec4} states the simulation results. Section \ref{sec5} gives the experimental results. Section \ref{sec6} summaries this paper and gives the future work.

\section{System model and problem formation}\label{sec2}
\subsection{System model}
\begin{figure}[h]
	\centering
	\includegraphics[width = 0.6\textwidth]{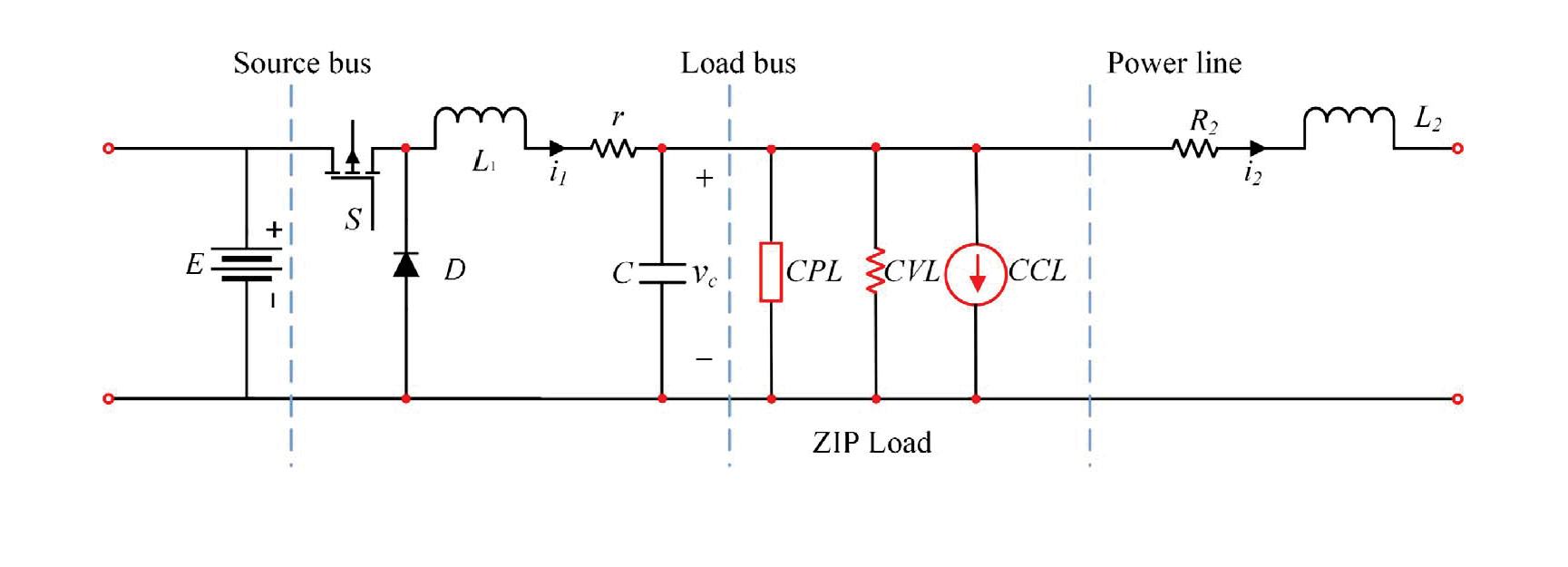}
\vspace{-0.2cm}
	\caption{A typical structure of buck converter based DC distribution system.}
	\label{Fig-ZIP load}
\end{figure}

Fig. \ref{Fig-ZIP load} shows a typical structure of DC distribution system, which is constructed by buck converter, DC source, ZIP load and power line.

Assuming that the converter operates in continuous conduction mode and borrowing the result in \cite{ferguson2020exponential}, its state space model of the system shown in Fig. \ref{Fig-ZIP load} is given by
\begin{equation}\label{model1}
\left\{\begin{aligned}
L^0_1 \dot{i}_{1}&={-r^0i_1}{+\mu E^0}-v_{c}+d_1, \\
C^0 \dot{v}_{c}&=i_{1}-\frac{v_{c}}{R^0}-\frac{P^0}{v_{c}}-i^0-i_2+d_2,\\
{L^0_2 \dot{i}_{2}}&={v}_{c}-R^0_2 i_2+d_3,
\end{aligned}\right.
\end{equation}
where {$v_{c}\in \mathbb{R}_{>0}$}, {$i_1\in \mathbb{R}, i_2 \in \mathbb{R}$} are capacitance voltage, inductor current, power line current.  In \eqref{model1}, $(\cdot)^0$ denotes nominal values of these circuit parameters. $R \in \mathbb{R}_{>0}$, $i \in \mathbb{R}_{>0}$ and $P \in \mathbb{R}_{>0}$ are resistance, current and power components of the ZIP load. {$r, R_2$ is the filter and power line resistances, respectively}. $E \in \mathbb{R}_{>0}$ denotes input voltage, $C \in \mathbb{R}_{>0}$ is capacitance, $L_1, L_2$ are inductances, $\mu \in [0, 1]$ is duty ratio. {$d_1, d_2, d_3$ are time-varyingly matched and mismatched disturbances, {where are given by
\begin{equation}
\left\{\begin{aligned}
d_1&=r^0 i_1+v_c-\mu E^0+\frac{L^0_1(-ri_1-v_c+\mu E)}{L_1},\nonumber\\
d_2&=-i_1+\frac{v_{c}}{R^0}+\frac{P^0}{v_{c}}+i^0+i_2+\frac{C^0(i_1-\frac{v_{c}}{R}-\frac{P}{v_{c}}-i-i_2)}{C}.\nonumber\\
d_3&=-{v}_{c}+R^0_2 i_2+\frac{L^0_2({v}_{c}-R_2 i_2)}{L_2}.\nonumber
\end{aligned}\right.
\end{equation}}
Note that the voltage of output port of the system shown in Fig. \ref{Fig-ZIP load} can be added to $d_3$. Here, we omit it. Hence, it is observed that $d_1=d_2=d_3=0$ when there is no parameter perturbation. Note that the construction of the lumped disturbances can be also found in \cite{chen2015disturbance,li2014disturbance}.}
Borrowing the results in \cite{chen2015disturbance,li2014disturbance}, it is supposed that the time-varying disturbances $d_i, i=1, 2, 3$ of the system \eqref{model1} are generated by the linear exogenous system
\begin{equation}
\left\{\begin{aligned}
\label{exo1}
\dot \zeta_i=&A_i \zeta_i,\\
d_i=&M_i\zeta_i,
\end{aligned}\right.
\end{equation}
where $\zeta_i \in \mathbb{R}^{m \times 1}$ is internal variable, $d_i \in \mathbb{R}$, $A_i\in \mathbb{R}^{m \times m}, M_i\in \mathbb{R}^{1 \times m}$ properly matrices. {Note that PH system captures symmetry and provides insights into the stability of the equilibrium and the system's phase space. Hence, to simplify the controller design and stability analysis, a PH form is adopted to describe the system.} {For convenient description, new states $x_1:=i_1$, $x_2:=v_c, x_3:=i_2$ are defined.} In the case,
the system (\ref{model1}) is rewritten as the PH form
{\begin{equation}\label{system2}
\begin{bmatrix}
\dot{{x}}_{1} \\
\dot{{x}}_{2} \\
\dot{{x}}_{3}
\end{bmatrix}=\begin{bmatrix}
{-\frac{r^0}{(L^0_1)^2}} & -\frac{1}{C^0 L_1^0} & 0 \\
\frac{1}{C^0 L_1^0} & -\frac{1}{R^0 (C^0)^{2}}-\frac{P^0}{x_{2}^{2} (C^0)^2} & -\frac{1}{C^0L^0_2}\\
0 & \frac{1}{C^0L^0_2} & -\frac{R^0_2}{(L^0_2)^2}
\end{bmatrix} \nabla H+\begin{bmatrix}
\frac{d_1}{L_1^0} \\
\frac{d_2}{C^0}-\frac{i^0}{C^0} \\
\frac{d_3}{L^0_2}
\end{bmatrix}+\begin{bmatrix}
\frac{\mu E^0}{L_1^0} \\
0\\
0
\end{bmatrix},
\end{equation}}
where the storage function is given by
{$
H=\frac{1}{2} L^0_1 x_{1}^{2}+\frac{1}{2} C^0 x_{2}^{2}+\frac{1}{2} L^0_2 x_{3}^{2},
$}
$\nabla$ denotes the transposed gradient.
For a given reference {constant} $x_2^\star=v_\star$, {reference dynamic system can be given as follows}
{\begin{equation}\label{expected dynamics}
	\left\{\begin{aligned}
		L^0_1 \dot{x}_{1}^{\star}&={-r^0{x}_{1}^{\star}}+\mu^\star E^0-x_{2}^{\star}+d_1, \\
		C^0 \dot{x}_{2}^{\star}&={x}_{1}^{\star}-\frac{{x}_{2}^{\star}}{R^0}-\frac{P^0}{{x}_{2}^{\star}}-i^0-{x}_{3}^{\star}+d_2,\\
		{L^0_2 \dot{x}_{3}^{\star}}&={x}_{2}^{\star}-R^0_2 {x}_{3}^{\star}+d_3.
	\end{aligned}\right.
\end{equation}}
Here we assume that the disturbances $d_1, d_2, d_3$ are measurable and they will be replaced by their estimates generated by the designed disturbance observer in Proposition \ref{pro2}.
{Because of $\dot{x}_{2}^\star=0,~\ddot{x}_{2}^\star=0$, the following equations can be obtained}
{\begin{equation}\label{equations for expected dynamics}
		\left\{\begin{aligned}
			 0&={x}_{1}^{\star}-\frac{{x}_{2}^{\star}}{R^0}-\frac{P^0}{{x}_{2}^{\star}}-i^0-{x}_{3}^{\star}+d_2,\\
			0&=\dot{x}_{1}^{\star}-\dot{x}_{3}^{\star}+\dot{d_2},\\
		\end{aligned}\right.
\end{equation}
By substituting the dynamic equations of $x_{1}$ and $x_{3}$ in (\ref{expected dynamics}) into (\ref{equations for expected dynamics}), {the reference dynamic of the system (\ref{system2}) satisfies}
\begin{equation}\label{equi5}
	\left\{\begin{aligned}
		\dot{x}_{1}^{\star}&=-\frac{r^0}{L^0_1} {x}_{1}^{\star}+\frac{\mu^\star E^0}{{L^0_1}}-\frac{x_{2}^{\star}}{L^0_1}+\frac{d_1}{L^0_1}, \\
		x_3^{\star}&=x_1^{\star}-\frac{x_2^{\star}}{R^0}-\frac{P^0}{x_2^*}-i^0+d_2,\\
		\mu^{\star}&=\frac{L_1^0}{E}\left(\frac{r^0 x_1^{\star}}{L_1^0}+\frac{x_2^{\star}}{L_1^0}+\frac{x_2^{\star}}{L_2^0}-\frac{R_2^0 x_3^{\star}}{L_2^0}-\frac{d_1}{L_1^0}+\frac{d_3}{L_2^0}-M_2 A_2 \zeta_2 \right),
	\end{aligned}\right.
	\end{equation}}

\subsection{Problem formulation}
The system \eqref{system2} verifies the following assumption.
\begin{assumption}\label{ass1}
The nominal parameters $L_1^0, E^0, r^0, C^0, R^0, P^0, i^0, L_2^0, R_2^0$ are known.
\end{assumption}

The control purpose is to regulate the state {$x_2$} around {$x_2^\star$}  in the presence of matched and mismatched disturbances $d_1, d_2, d_3$.
{A domain of attraction $\Omega$ is defined, where the set $\Omega$ inside the first quadrant verifies
{\begin{align}
(x_1(0), x_2(0), x_3(0)) \in \Omega &\Rightarrow (x_1(t), x_2(t), x_3(t)) \in \Omega,\nonumber\\
	\lim\limits_{t \rightarrow \infty} (x_1(t), x_2(t), x_3(t))&=(x^\star_1, x_2^\star, x^\star_3)
\end{align}}}
\section{Adaptive ESC design and stability analysis}\label{sec3}
The controller is proposed by following three steps, which are given by
\begin{enumerate}
	\item[(1)] Assuming that the disturbances $d_1, d_2, d_3$ are known, an ESC is designed to ensure that {$x_2$} tracks desired value {$x_2^\star$}.
	\item[(2)] A disturbance observer is designed to online estimate the disturbances $d_1, d_2, d_3$. The estimation $\hat d_1, \hat d_2, \hat d_3$ generated by it can asymptotically converge to their actual values.
	\item[(3)] Combining the ESC with the disturbance observer is to form an AESC, which ensures that the closed-loop system is asymptotically stable.
\end{enumerate}
\subsection{ESC design}
{For a given $x_2^\star$, the errors are defined as $e_1:=x_1-x_1^\star,~e_2:=x_2-x_2^\star,~e_3:=x_3-x_3^\star$. Using \eqref{model1} and \eqref{expected dynamics}, the error dynamics can be written as
\begin{equation}\label{system3}
\begin{bmatrix}
\dot{{e}}_{1} \\
\dot{{e}}_{2} \\
\dot{{e}}_{3}
\end{bmatrix}=\begin{bmatrix}
{-\frac{r^0}{(L_1^0)^2}} & -\frac{1}{C^0 L_1^0} & 0\\
\frac{1}{C^0 L_1^0} & -\frac{1}{R^0 (C^0)^{2}}+\frac{P^0}{x_{2} x_2^{*} (C^0)^{2}} &-\frac{1}{C^0L^0_2}\\
0 & \frac{1}{C^0L^0_2}& -\frac{R^0_2}{(L^0_2)^2}
\end{bmatrix} \nabla {H}_1+\begin{bmatrix}
u \\
0\\
0
\end{bmatrix}
\end{equation}
with storage function
$
H_1=\frac{1}{2} L^0_1 e_{1}^{2}+\frac{1}{2} C^0 e_{2}^{2}+\frac{1}{2} L^0_2 e_{3}^{2},
$
where the relationship between $\mu$ and a new control signal $u$ is given by
\begin{equation}\label{transform}
\mu=\frac{L_1^0u}{E^0}+\mu^{\star}.
\end{equation}}
\begin{proposition}\label{controllerp}
	Consider the system (\ref{system3}) verifying Assumption \ref{ass1}. An ESC is designed as {\begin{equation}
\left\{\begin{aligned}
	\dot{x}_{c}&=-\alpha e_2, \\
	u&=-\alpha r^0 k\left(\alpha e_{1}-\frac{x_{c}}{L_1^0}\right), \label{controller}
\end{aligned}\right.
\end{equation}}
	where $k > 0$ and $\alpha > 0$ are controller gains, $x_c$ is a new state of the controller. The following condition is satisfied
	\begin{equation}\label{condition}
	\frac{R^0 P^0}{x_{2} x_{2}^{*}}<1.
	\end{equation}
	Then, the closed-loop system is asymptotically stable around the equilibrium {$(e_1^\star, e_2^\star ,e_3^\star)=(0,0,0)$}.
\end{proposition}
{\it Proof:}
Define state vector as {$ \boldsymbol{e}:=(e_1, e_2, e_3, x_c)^T$}. The new storage function is chosen as
{\begin{equation}\label{lypno}
H_d(\boldsymbol{e})=\frac{1}{2} L^0_1 e_1^{2}+\frac{1}{2} C^0 e_2^{2}+\frac{1}{2} L^0_2 e_3^{2}+\frac{k}{2}(\alpha L_1^0 e_{1}-x_{c})^{2}.
\end{equation}}
It is observed that the storage function $H_d$ given by \eqref{lypno} takes the minimum value at the desired equilibrium {$(e_1^\star, e_2^\star ,e_3^\star, x_c^\star)=(0,0,0,0)$}.
Substituting the controller \eqref{controller} in the system \eqref{system3} obtains the closed-loop system
{\begin{equation}\label{closed-loop}
\dot{ {\boldsymbol{e}}}=\begin{bmatrix}
-\frac{r^0}{(L_1^0)^{2}} & -\frac{1}{C^0 L_1^0} & 0 & 0 \\
\frac{1}{C^0 L_1^0} & -\frac{1}{R^0 (C^0)^{2}}+\frac{P^0}{x_{2} x_2^{*} (C^0)^{2}} & -\frac{1}{C^0L^0_2} & \frac{\alpha}{C^0} \\
0 & \frac{1}{C^0L^0_2} & -\frac{R^0_2}{(L^0_2)^2} & 0\\
0 & -\frac{\alpha}{C^0} & 0 & 0
\end{bmatrix} \nabla H_d.
\end{equation}
The derivative of the storage function with respect to time is given by
\begin{equation}\label{The derivative of the storage function}
\dot{H_d}(\boldsymbol{e})=\nabla^{T} H_d( \boldsymbol{e}) \dot{\boldsymbol{e}}.
\end{equation}
Substituting (\ref{closed-loop}) into (\ref{The derivative of the storage function}), it obtains
\begin{equation}\label{tdh}
\begin{aligned}
\dot{H_d}(\boldsymbol{e}) =\nabla^{T} H_d(\boldsymbol{e})\begin{bmatrix}
-\frac{r^0}{(L_1^0)^2} & 0 & 0 & 0\\
0 & -\frac{1}{R^0 (C^0)^{2}}+\frac{P^0}{x_{2} x_2^{*} (C^0)^{2}} & 0 & 0 \\
0 & 0 & -\frac{R^0_2}{(L^0_2)^2} &0 \\
0 & 0 & 0 &0
\end{bmatrix} \nabla H_d( \boldsymbol{e}).
\end{aligned}
\end{equation}
If condition (\ref{condition}) are satisfied, the function $\dot{H}_d(\boldsymbol{e})$ is negative semi-definite. We define a set {$\Phi:=\{ \boldsymbol{e}|~\dot{H}_d( \boldsymbol{e})=0\}$.} From \eqref{lypno}, the largest invariant set contained in the set {$\Phi$} can be computed as $\{ \boldsymbol{e}|(0, 0, 0, 0)\}$. Therefore, the system \eqref{closed-loop} is asymptotically stable and state $\boldsymbol{e}$ can converge to the equilibrium. The proof is completed.}

It follows \eqref{transform} that the ESC is given by
{\begin{equation}\label{controller2}
\mu=\frac{\alpha r^0 k}{E^0} \left[ x_{c}-\alpha L_1^0 (x_1-x_1^\star)\right]+\mu^{\star}.
\end{equation}}
It is noted that sufficient condition (\ref{condition}) depends on the trajectory of state $x_2$, which is difficult to verify for practical applications. In the following subsection, the convergence condition (\ref{condition}) will be further investigated.

{\subsection{Further discussion on sufficient condition \eqref{condition}}}
Next, for the condition (\ref{condition}), an alternative one relating with initial condition is given to ensure the convergence.

Firstly, the condition \eqref{condition} is given by a set ${N}=\{\boldsymbol {x}|~x_2>\frac{R^0P^0}{x_2^*}\}$. Since $x_2>0$, one has  ${N}=\{\boldsymbol {x}|~x_2^*-x_2<x_2^*-\frac{R^0P^0}{x_2^*}\}.$ Then, a subset $B_1$ of set ${N}$ is obtained as
$
B_1 = \{ \boldsymbol{x}| ~|x_2^*-x_2|<x_2^*-\frac{R^0P^0}{x_2^*}\}=\{ \boldsymbol{e}| ~|e_2|<x_2^*-\frac{R^0P^0}{x_2^*}\}.
$
Furthermore, the set $B_1$ can be given by
$
B_1=\{ \boldsymbol{e}| ~\sqrt{\frac{2}{C^0} H_{d}(\boldsymbol{e})}<x_2^*-\frac{R^0P^0}{x_2^*}+\sqrt{\frac{2}{C^0} H_{d}(\boldsymbol{e})}-|e_2|\}.
$
According to the storage function \eqref{lypno}, we can get $\sqrt{\frac{2}{C^0} H_{d}(\boldsymbol{e})}-|e_2|>0$. Here, we define a set $B_2=\{ \boldsymbol{e}|~\sqrt{\frac{2}{C^0} H_{d}(\boldsymbol{e})}<x_2^*-\frac{R^0P^0}{x_2^*}\}$, which is a subset of set $B_1$ and also a subset of set {$N$}. In the case, a new condition is given by
\begin{equation}\label{c2}
\sqrt{\frac{2}{C^0} H_{d}(\boldsymbol{e})}<x_2^*-\frac{R^0P^0}{x_2^*}.
\end{equation}
Next, the condition (\ref{c2}) will be generalized into the form which is only dependent on the initial state $\boldsymbol{e}_{0}$.

\begin{proposition}\label{rewritep2}
	For $t \geq 0 $, if initial state $\boldsymbol{e}_0$ satisfies $\sqrt{\frac{2}{C^0} H_{d}\left(\boldsymbol{e}_{0}\right)}<x_{2}^{*}-\frac{R^0P^0}{x_{2}^{*}}$, then the inequality $\sqrt{\frac{2}{C^0} H_{d}\left(\boldsymbol{e}\right)}<x_{2}^{*}-\frac{R^0P^0}{x_{2}^{*}}$ holds.
\end{proposition}

{\it Proof:}
Firstly, time is discretized into infinite number of time points with infinitesimal intervals. In order to facilitate the description, a strictly increasing sequence $\{t_n\}$  with $n \in \mathbb{Z}_{\geq 0}$ is defined to represent the discretized time. The elements of the sequence $\{t_n\}$ correspond to the discrete time one by one. Sequence $\{t_n\}$ has the following characteristics
\begin{enumerate}
	\item[1)] The first term of the sequence is equal to 0.
	\begin{equation*}
	t_0=0.
	\end{equation*}
	
	\item[2)] The values of adjacent elements are infinitely close.
	\begin{equation*}
	t_{n+1}-t_{n} \rightarrow 0, n \in \mathbb{Z}_{\geq 0}.
	\end{equation*}
\end{enumerate}
At $t=t_n$, the value of state $\boldsymbol{e}$ is denoted by $\boldsymbol{e}_n$. Then, Proposition \ref{rewritep2} can be described in another form. For $ \forall n\in \mathbb{Z}_{\geq 0}$, if  $\sqrt{\frac{2}{C^0} H_{d}\left(\boldsymbol{e}_{0}\right)}<x_{2}^{*}-\frac{R^0P^0}{x_{2}^{*}}$, then inequality $\sqrt{\frac{2}{C^0} H_{d}\left(\boldsymbol{e}_n\right)}<x_{2}^{*}-\frac{R^0P^0}{x_{2}^{*}}$ holds.

Next, the above proposition will be proved by mathematical induction method.
\begin{enumerate}
	\item[(i)] Considering $n=0$, it is obvious that inequality
	$
	\sqrt{\frac{2}{C^0} H_{d}\left(\boldsymbol{e}_0\right)}<x_{2}^{*}-\frac{R^0P^0}{x_{2}^{*}}
	$
	is correct.
	\item[(ii)] Assuming that
	\begin{equation}\label{tk}
	\sqrt{\frac{2}{C^0} H_{d}\left(\boldsymbol{e}_k\right)}<x_{2}^{*}-\frac{R^0P^0}{x_{2}^{*}}
	\end{equation}
	with $k\in \mathbb{Z}_{\geq 0}$ is true, the existence of inequality
	\begin{equation}\label{tk+1}
	\sqrt{\frac{2}{C^0} H_{d}\left(\boldsymbol{e}_{k+1}\right)}<x_{2}^{*}-\frac{R^0P^0}{x_{2}^{*}}
	\end{equation}
	will be verified.
\end{enumerate}
	
	\begin{equation*}
	\begin{aligned}
	\sqrt{\frac{2}{C^0} H_{d}(\boldsymbol{e}_k)}&<x_{2}^{*}-\frac{R^0P^0}{x_{2}^{*}},\\
	|e_2(t_k)|&<x_{2}^{*}-\frac{R^0P^0}{x_{2}^{*}} \Leftarrow [e_2(t_k)]^2 \leq \frac{2}{C^0} H_{d}(\boldsymbol{e}_k),\\
	|x_2(t_k)-x_2^*|&<x_{2}^{*}-\frac{R^0P^0}{x_{2}^{*}},\\
	x_2^*-x_2(t_k)&<x_{2}^{*}-\frac{R^0P^0}{x_{2}^{*}}\Leftarrow x_2^*-x_2(t_k) \leq |x_2^*-x_2(t_k)|,\\
	x_2(t_k)&>\frac{R^0P^0}{x_{2}^{*}}.
	\end{aligned}
	\end{equation*}
\begin{enumerate}	
	\item[]From (\ref{tdh}), we can get $H_d(\boldsymbol{e}_{k+1}) \leq H_d(\boldsymbol{e}_k)$. Then, according to (\ref{tk}), the inequality (\ref{tk+1}) holds.
\end{enumerate}
Combining (i) with (ii), the proof of Proposition \ref{rewritep2}  is completed.

Consequently, the new condition can be expressed as
\begin{equation}\label{nc}
\sqrt{\frac{2}{C^0} H_{d}(\boldsymbol{e}_0)}<x_2^*-\frac{R^0P^0}{x_2^*}.
\end{equation}
Note that condition (\ref{nc}) depends on initial state $\boldsymbol{e}_{0}$, which can be freely defined in practical applications. By choosing  appropriate initial condition $\boldsymbol{e}_{0}$ satisfying condition (\ref{nc}), a reliable controller is obtained with the ensured convergence. The domain defined in \eqref{nc} will be verified in Section \ref{sec4} by simulation study.

\begin{remark}
	The controller contains an integral action around tracking error $x_2-x_{2\star}$. This will eliminate the affect of the unmodeled dynamics and other parameter perturbations on the system.
\end{remark}
\subsection{Adaptive ESC design}
In this subsection, a disturbance observer is designed for the system (\ref{model1}).
{The construction of the following disturbance observer adopts the result in \cite[Section 3.3.2]{li2014disturbance}.}

{
\begin{proposition}\label{pro2}
For the system \eqref{model1} verifying the conditions \eqref{exo1}, a disturbance observer is designed as
\begin{equation}
\left\{
\begin{aligned}\label{bucest2}
    \hat d_1=&M_1\hat \zeta_1,
    \hat d_2=M_2\hat \zeta_2,
    \hat d_3=M_3\hat \zeta_3,\\
\hat \zeta_1=& z_1{+L^0_1p_1(x_1)},
\hat \zeta_2= z_2{+C^0p_2(x_2)},
\hat \zeta_3= z_3{+L^0_2p_3(x_3)},\\
  \dot {z}_1=&(A_1-l_1({x_1})M_1)z_1+A_1 L_1^0p_1({x_1})-l_1({x_1})(M_1L^0_1p_1({x_1})
\\&-r^0x_1-x_2+\mu E^0),\\
\dot {z}_2=&(A_2-l_2({x_2})M_2)z_2+A_2 C^0p_2({x_2})-l_2({x_2})(M_2C^0p_2({x_2})\\&+x_1-\frac{x_{2}}{R^0}-\frac{P^0}{x_{2}}-i^0-x_3),\\
\dot {z}_3=&(A_3-l_3({x_3})M_3)z_3+A_3 L^0_2p_3({x_3})-l_3({x_3})(M_3L^0_2p_3({x_3})\\&+x_2-R_2^0x_3),
	\end{aligned}
\right.
\end{equation}
where $z_i \in \mathbb{R}^{m \times 1}$ is the internal state variables of the observer and
$p_i({x_i}) \in \mathbb{R}^{m \times 1}$ is a nonlinear function to be designed. The nonlinear observer gain $l_i({x_i})$ is then determined by $l_i(x)=\frac{\partial p_i({x_i})}{\partial {x_i}}.$
The following condition holds.
\begin{align}\label{d1d2}
A_1-l_1(x_1)M_1<0, A_2-l_2(x_2)M_2<0, A_3-l_3(x_3)M_3<0.
\end{align}
Then, $\hat \zeta_i$ approaches to $\zeta_i$ asymptotically, which is stated as
\begin{align}\label{esti}
    \lim_{t \to \infty}\tilde \zeta_1(t)=0,
\lim_{t \to \infty}\tilde \zeta_2(t)=0,
\lim_{t \to \infty}\tilde \zeta_3(t)=0
	\end{align}
with $\tilde \zeta_1(t)=\hat \zeta_1(t)-\zeta_1(t), \tilde \zeta_2(t)=\hat \zeta_2(t)-\zeta_2(t), \tilde \zeta_3(t)=\hat \zeta_3(t)-\zeta_3(t)$.
Therefore, the observer \eqref{bucest2} can asymptotically estimate the disturbances.
\end{proposition}
}

\textit{Proof:}
Defining the errors
 $\tilde \zeta_1:=\hat \zeta_1-\zeta_1, \tilde \zeta_2:=\hat \zeta_2-\zeta_2, {\tilde \zeta_3:=\hat \zeta_3-\zeta_3}$
 and differentiating them along the trajectory \eqref{model1} have
$
\dot {\tilde \zeta}_1=(A_1-l_1(x_1)M_1)\tilde \zeta_1,
\dot {\tilde \zeta}_2=(A_2-l_2(x_2)M_2)\tilde \zeta_2,
\dot {\tilde \zeta}_3=(A_3-l_3(x_3)M_3)\tilde \zeta_3.
$
Therefore, we can choose $l_1, l_2, l_3$ such that the condition \eqref{d1d2} holds. Then, the errors $\tilde \zeta_1, \tilde \zeta_2, \tilde \zeta_3$ asymptotically converge to zero, which implies that the observer \eqref{bucest2} can asymptotically estimate the disturbances. The detailed analysis is shown in \cite[Section 3.3.2]{li2014disturbance}.

Introducing  the estimate terms $\hat d_1, \hat d_2, \hat d_3, \hat{\zeta}_2$ into the controller (\ref{controller2}) and the reference dynamics \eqref{expected dynamics} and borrowing \eqref{equi5}  yields the AESC
{\begin{equation}
\left\{
\begin{aligned}
\label{controller3}
\hat \mu&=\frac{\alpha r^0 k}{E^0} \left[ x_{c}-\alpha L_1^0 (x_1-\hat{x}_1^\star)\right]+\hat{\mu}^{\star},\\
\hat{\mu}^{\star}&=\frac{L_1^0}{E}\left(\frac{r^0 \hat{x}_1^{\star}}{L_1^0}+\frac{x_2^{\star}}{L_1^0}+\frac{x_2^{\star}}{L_2^0}-\frac{R_2^0 \hat{x}_3^{\star}}{L_2^0}-\frac{\hat{d}_1}{L_1^0}+\frac{\hat{d}_3}{L_2^0}-M_2 A_2 \hat{\zeta}_2 \right),\\
\dot{\hat{x}}_{1}^{\star}&=-\frac{r^0}{L^0_1} \hat{x}_{1}^{\star}+\frac{\hat{\mu}^\star E^0}{{L^0_1}}-\frac{x_{2}^{\star}}{L^0_1}+\frac{\hat{d}_1}{L^0_1}, \\
\hat{x}_3^{\star}&=\hat{x}_1^{\star}-\frac{x_2^{\star}}{R^0}-\frac{P^0}{x_2^*}-i^0+\hat{d}_2,\\
\end{aligned}\right.
\end{equation}}

{
\subsection{Stability analysis}
Here, the stability analysis of the system  under the AESC (\ref{controller3}) will be investigated. The control law $\hat{\mu}$ can be expressed in a perturbed form
\begin{equation}
\hat{\mu}=\mu+\delta_{1}(\tilde{\zeta})
\end{equation}
and
$
\dot{\hat x}^\star_1=\dot{x}^\star_1+\delta_{2}(\tilde{\zeta}),
\hat x^\star_3=x^\star_3+\delta_{3}(\tilde{\zeta}),
$
where $\delta_1 (\tilde{\zeta}), \delta_{2}(\tilde{\zeta}), \delta_{3}(\tilde{\zeta})$ are perturbation terms with $\tilde{\zeta}=[\tilde{\zeta}_1~ \tilde{\zeta}_2~\tilde{\zeta}_3]^T$.
Recalling \eqref{closed-loop}, the cascaded closed-loop system can be represented as
{\begin{equation}\label{system4}
\left\{\begin{aligned}
\dot{ \boldsymbol{e}}&=\begin{bmatrix}
-\frac{r^0}{(L_1^0)^{2}} & -\frac{1}{C^0 L_1^0} & 0 & 0 \\
\frac{1}{C^0 L_1^0} & -\frac{1}{R^0 (C^0)^{2}}+\frac{P^0}{x_{2} x_2^{*} (C^0)^{2}} & -\frac{1}{C^0L^0_2} & \frac{\alpha}{C^0} \\
0 & \frac{1}{C^0L^0_2} & -\frac{R^0_2}{(L^0_2)^2} & 0\\
0 & -\frac{\alpha}{C^0} & 0 & 0
\end{bmatrix} \nabla H_d+\delta (\tilde{\zeta}),\\
\dot{\tilde{\zeta}}&=\Psi
\tilde{\zeta},
\end{aligned}\right.
\end{equation}}
where $\Psi$ is a properly defined matrix, $\delta (\tilde{\zeta}) \in \mathbb{R}^{4}$ is a perturbed term constructed by $\delta_1, \delta_{2}, \delta_{3}$ and $\tilde{\zeta}$.  Note that $\delta(0)=0$.
Based on Proposition \ref{controllerp}, it can be deduced that if $\delta=0$, the closed-system is asymptotically stable under the condition (\ref{nc}). From Proposition \ref{pro2}, the error system--$\tilde{\zeta}$ is asymptotically stable with the condition \eqref{d1d2}. Using the Proposition 4.1 in \cite{sepulchre2012constructive} related with the stability of cascaded system, the locally asymptotic stability of overall system (\ref{system4}) is established.
}

{
In summary, the design steps of the proposed controller and the qualitative ideas behind the selection of controller and
observer parameters are stated as follows.
\begin{enumerate}
	\item[(1)] It is assumed that the disturbances $d_1, d_2, d_3$ are known. Designing the ESC  (\ref{controller}) with $\alpha > 0, k > 0$ is to ensure that the closed-loop system is asymptotically stable. Increasing the parameters $\alpha, k$ can decrease the convergence time. However, large parameters may cause big noise so that the tradeoff between them should be considered.
	\item[(2)] The disturbance observer is developed with the gains $l_1(x_1), l_2(x_2), l_3(x_3)$ to estimate the matched and mismatched disturbances in real time. The convergence of the observer is accelerated by increasing the gains. Similarly, the tradeoff between transient and noise level should be noted.
	\item[(3)]	Replacing $d_1, d_2, d_3$ with their estimates $\hat{d}_1, \hat{d}_2, \hat{d}_3$ in the reference dynamic \eqref{expected dynamics} and the controller \eqref{transform} and (\ref{controller})  is to construct an AESC \eqref{controller3}.
\end{enumerate}
}
{
\begin{remark}
Compared with the existing results \cite{hassan2018adaptive,9646879,he2022adaptive}, several main differences are stated as follows.
\begin{itemize}
  \item In \cite{hassan2018adaptive,9646879}, the authors propose  some excellent control methods to deal with the control problem under the CPL.
  \item {In \cite{he2022adaptive}, an adaptive sensorless controller is proposed for buck converter with CPL. It is difficult to expect that this result can deal with the control problem of buck converter with ZIP load based DC distribution system since the excitation condition can not be satisfied in practical application. Hence, the reconstruction of ZIP load and states is hindered.}
\end{itemize}
\end{remark}
}
{
\begin{remark}
Unsurprisingly, there are some methods for dealing with the control problem of buck converter with ZIP load. A qualitative discussion on the controller complexity of several schemes are given as follows.
\begin{itemize}
  \item It is observed from that an ESC is designed to stabilize the system in this paper, which considers the presence of time-varying disturbances.
  \item Integrating ESC with disturbance observer is to form an AESC approach with a relatively hight complexity, which is to mainly improve the robustness.
  \item A backstepping control scheme can be proposed for the system, which may be more complicated than the designed ESC.
  \item A model predictive control can be also presented, but its computational load of CPU should be noticed.
\end{itemize}
\end{remark}
}
{
\begin{remark}
It is noted that we consider a ZIP load of buck converter based DC distribution system in this paper, which contains a CPL. For the mathematical difficulty of handling CPL, the issues are from the dynamic and nonlinear characteristics of CPL within power system. Here are some of the key mathematical difficulties considered:
\begin{itemize}
  \item Stability analysis: CPL can significantly affect the stability of power system, such as buck converter based DC distribution system. Due to its nature of maintaining a constant power flow, variations in voltage lead to inverse variations in current, which can create negative impedance characteristics. Mathematically modeling these effects to predict and analyze stability, especially under different operating conditions, requires employing dynamic system analysis techniques, which is held in the proposed method of this paper.
  \item Control system design: Integrating CPL into power system (like buck converter based DC distribution system) necessitates sophisticated control system to manage the power flow and maintain system stability. Designing the controller involves implementing control theory, often requiring complex mathematical tools such as nonlinear control techniques like energy shaping, which is adopted in our design.
  \item Simulation and experiment of transient responses: Transient analysis becomes critical when CPL is a part of a system since their response to changes in system conditions (like voltage sags or swells or power steps and parameter variations) can lead to significant instabilities. Simulation and experimentation of these transient responses to predict and mitigate potential problems involves the proposed control method.
  \item Interactions with other system components: The interaction of CPL with other dynamic components in a power system, such as input voltage and power loads, complicates the overall system behavior. Analyzing these interactions often requires the use of system observing methods, which is carried out in our paper.
\end{itemize}
Addressing these difficulties are considered using the proposed energy shaping control method with various trials of both theory and application.
\end{remark}
}
\section{Simulation results}\label{sec4}
\begin{figure}
	\centering
	\includegraphics[width =0.5\textwidth]{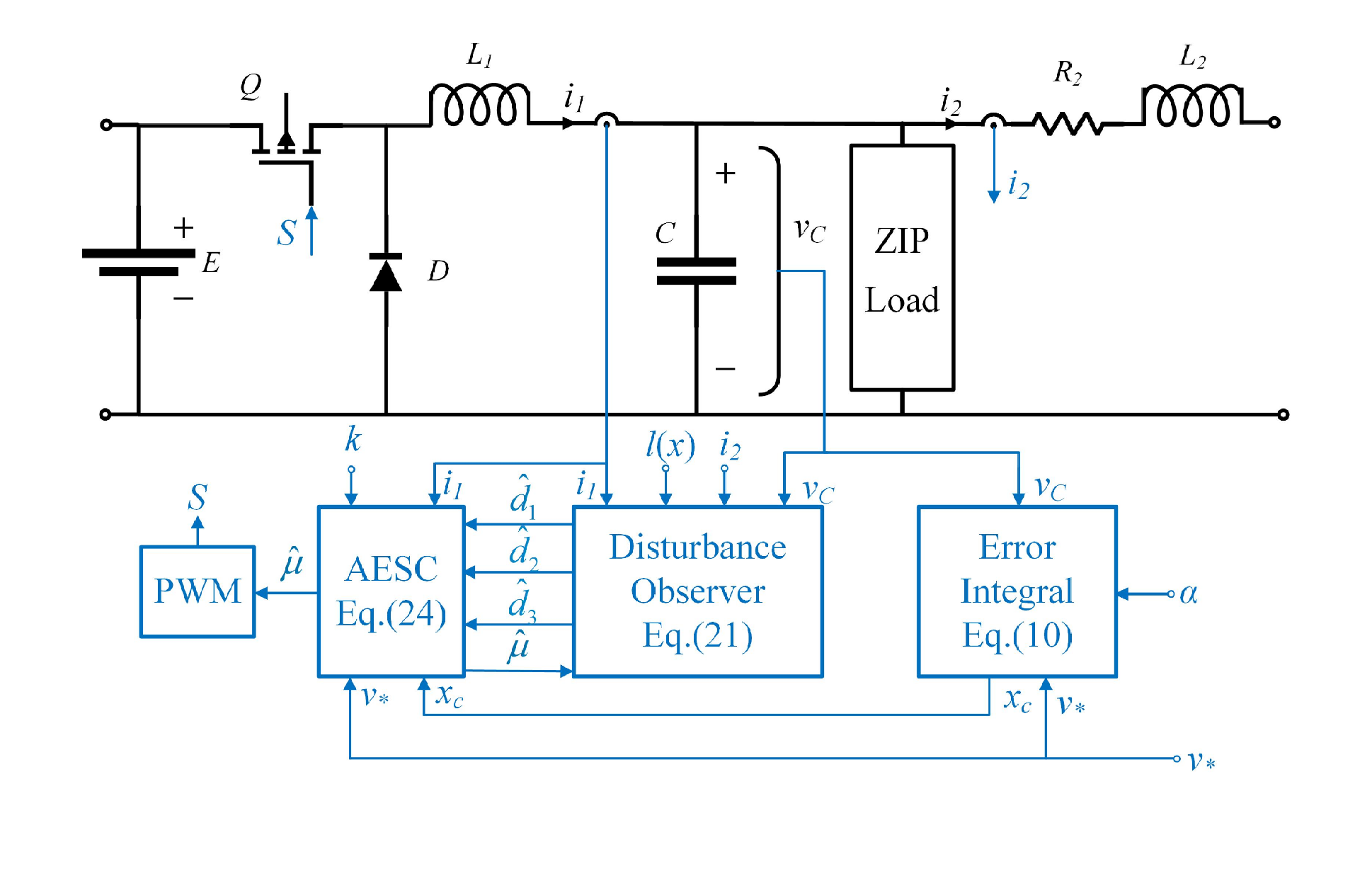}
\vspace{-0.5cm}
	\caption{{The control diagram of DC-DC buck converter feeding a ZIP load.}}
	\label{Fig:control}
\end{figure}

Borrowing the software Matlab/Simulink, the performance of the control algorithm is assessed. The AESC diagram is shown in Fig. \ref{Fig:control}. Tab. \ref{t1} reveals nominal circuit parameters of the system. {
The initial condition is selected as $i_1(0)=6$ A, $v_c(0)=15$ V, $i_2(0)=1$ A, $x_c(0)=-1$.
}

\begin{table}
	\caption{{Nominal circuit parameters}\label{t1}}
	\centering
	\begin{tabular}{lcc}
		\hline
		\hline
		Parameters & Symbols & Values\\
		\hline
		Input voltage & $E$ & 30 V\\
		Inductance & $L_1$ & 110 $\mu$H\\
        Inductance & $L_2$ & 110 $\mu$H\\
		Capacitance & $C$ & 1200 $\mu$F\\
        Parasitic resistance & $r$ & 0.15 $\Omega$\\
		Current load component & $i$ & 1 A\\
		Power load component & $P$ & 20 W\\
		Resistance load component & $ {R}$ & 5 $\Omega$\\
        Resistance & $R_2$ & 20 $\Omega$\\
		Reference  & $v_*$ & 20 V\\
		\hline
		\hline
	\end{tabular}
\end{table}

\begin{figure}
	\centering
	\includegraphics[scale=0.6]{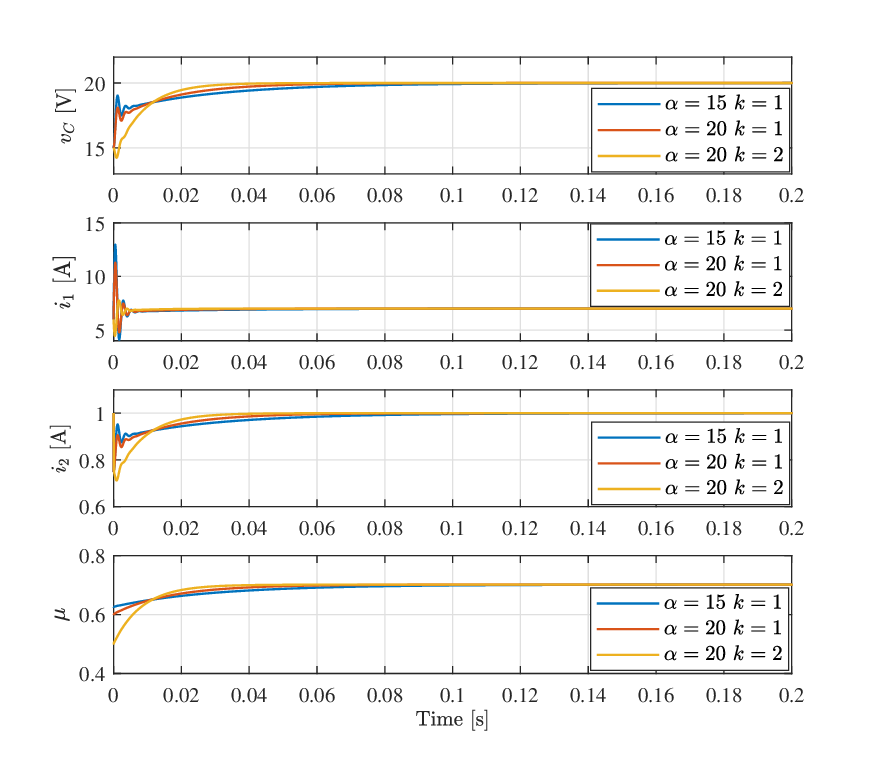}
\vspace{-0.3cm}
	\caption{The response curves of buck converter with a ZIP load under the proposed controller with different control gains.}
	\label{Fig:simulation 1}
\end{figure}

\begin{figure}
	% \vspace{-0.5cm}
	\centering
	\includegraphics[scale=0.6]{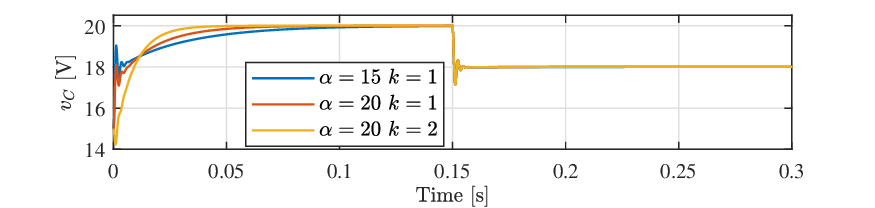}
	\caption{{Output voltage curves in the case of a step change in the reference $v_*$.}}
	\label{Fig:simulation 2}
\end{figure}

{{\bf Case 1}
Firstly, we test the performance of controller (\ref{controller3}) under different controller gains. As can be seen in Fig. \ref{Fig:simulation 1}, speed of convergence of the system states increases as the gains $\alpha$, $k$ are increased. Fig. \ref{Fig:simulation 2} displays the output curve in the presence of a step change in the reference $v_*$. The reference $v_*$ is varied from 20 V to 18 V at 0.15 s. The tracking error of the output voltage can converge to zero quickly in the presence of a step change in the reference $v_*$.}

\begin{figure}
	\centering
	\includegraphics[scale=0.6]{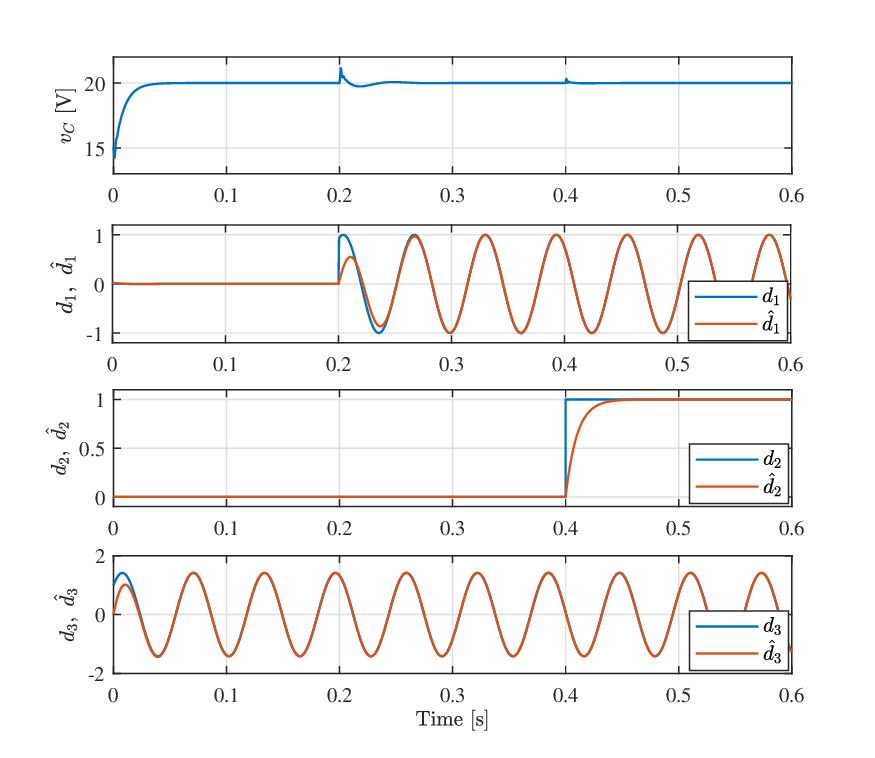}
\vspace{-0.3cm}
	\caption{{Curves of output voltage and observer output  when perturbations are introduced.}}\label{Fig:simulation 3}
\end{figure}

{{\bf Case 2}
Next, The effect of perturbations on closed-loop systems under the controller (\ref{controller3}) is investigated. The disturbances $d_3=\mathrm{sin}(100t)+\mathrm{cos}(100t)$, $d_1=\mathrm{sin}(100 t
)$,$d_2=1$ are introduced at 0 s, 0.2 s, 0.4 s respectively, which is shown in Fig. \ref{Fig:simulation 3}. {This implies that $M_1=[1~0], M_2=1, M_3=[1~1]$ and
$
A_1=A_3=\left[
      \begin{array}{cc}
        0 & 100 \\
        -100 & 0 \\
      \end{array}
    \right],
A_2=0.
 $}
 The observer parameters are selected as $p_1(i_1)=(99i_1,20i_1)^{T}$, $p_2(v_c)=100v_c$, $p_3(i_2)=(100i_2,100i_2)^{T}$. Fig. 6 shows that the output of observer  can quickly track the actual value of perturbations. The perturbations $d_1, d_2, d_3$ cause only small short-term fluctuations in the output voltage.}
\begin{figure}
	\includegraphics[scale=0.62,trim=-200 0 0 0,clip]{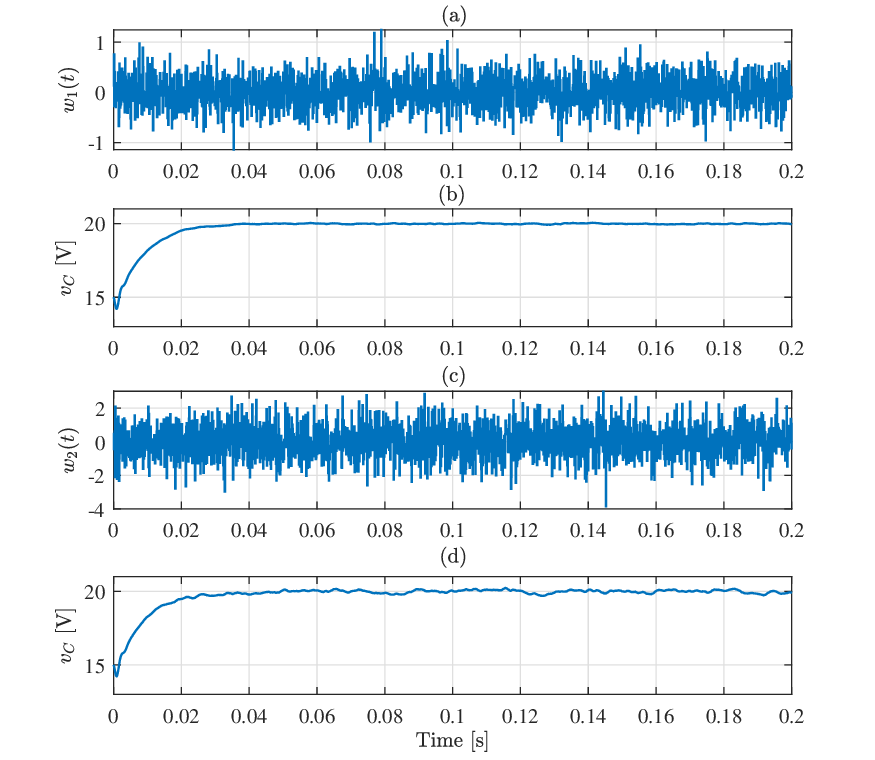}
\vspace{-0.1cm}
	\caption{{Curves of measurement noise and output voltage. (a) White noise $w_1(t)$ with low power, (b) output voltage $v_c$ in the presence of measurement noise $w_1(t)$}, (c) white noise $w_2(t)$ with high power, (d) output voltage $v_c$ in the presence of measurement noise $w_2(t)$. }\label{Fig:noise}
\end{figure}
\begin{figure}
	\centering
	\includegraphics[scale=0.6]{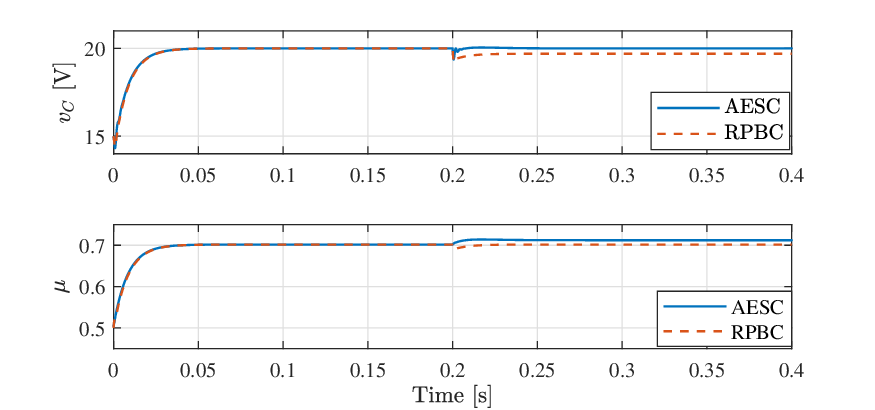}
	\caption{The response curves of the system with both controllers.}
	\label{Fig:simulation 4}
\end{figure}

\begin{figure}[!ht]
\begin{minipage}[t]{0.49\linewidth}
\centering
\includegraphics[width=0.65\textwidth]{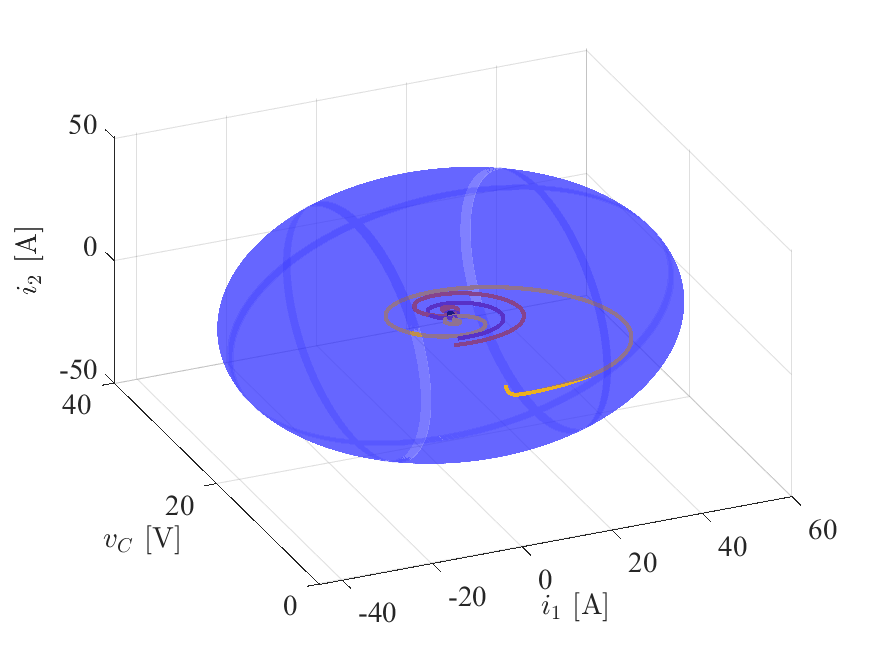}
\centerline{\tiny{(a)}}
\end{minipage}
\begin{minipage}[t]{0.49\linewidth}
\centering
\includegraphics[width=0.65\textwidth]{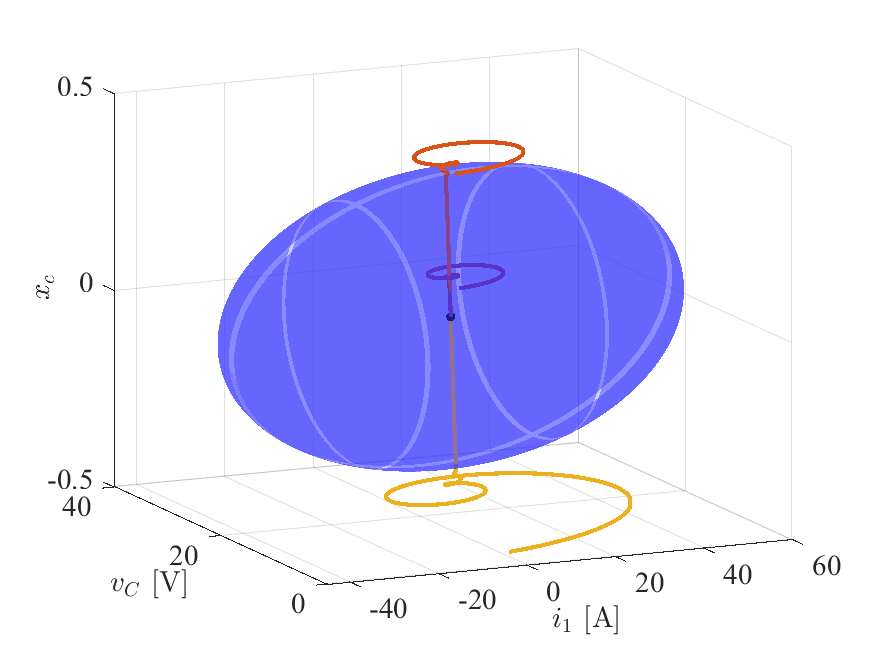}
\centerline{\tiny{(b)}}
\end{minipage}
\begin{minipage}[t]{0.49\linewidth}
\centering
\includegraphics[width=0.65\textwidth]{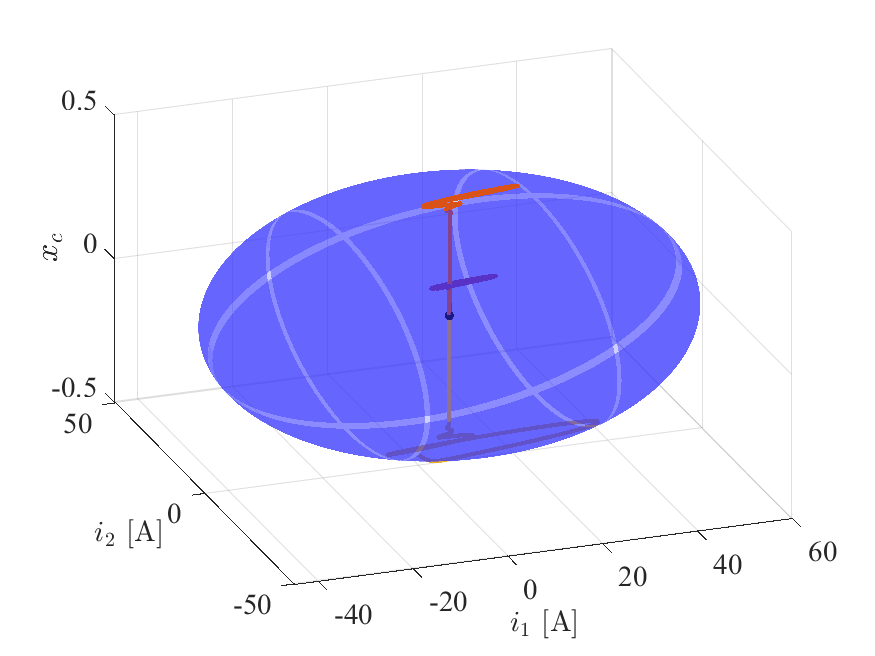}
\centerline{\tiny{(c)}}
\end{minipage}
\begin{minipage}[t]{0.49\linewidth}
\centering
\includegraphics[width=0.65\textwidth]{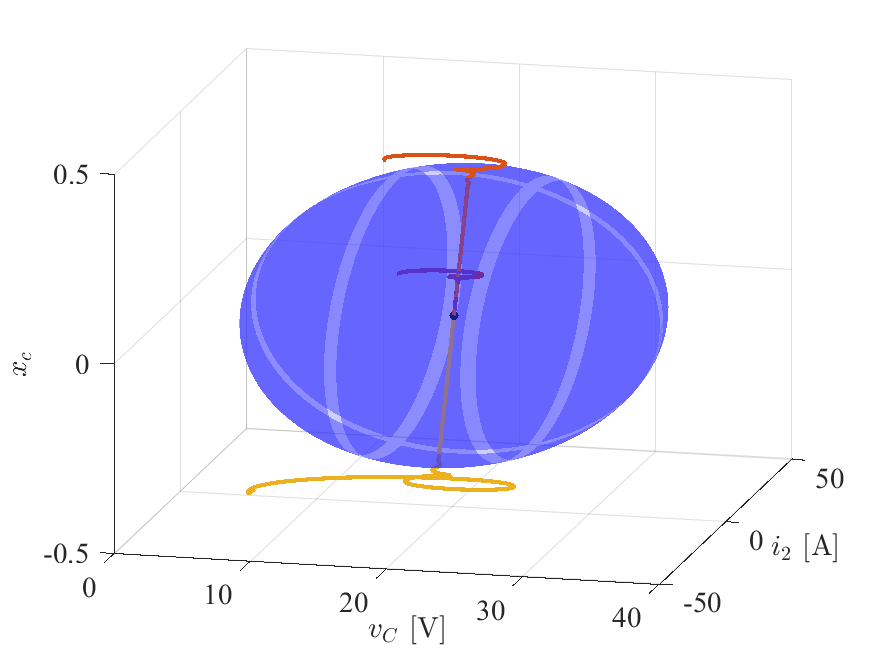}
\centerline{\tiny{(d)}}
\end{minipage}
\caption{Three-dimensional projections of convergence domains and state trajectories~ (a)~$i_1,~v_c,~i_2$;~(b)~$i_1,~v_c,~x_c$;~ (c)~$i_1,~i_2,~x_c$;~ (d)~$v_c,~i_2,~x_c$.}\label{fig3}
\end{figure}

{
 {\bf Case 3}
Subsequently, we will test the robustness against measurement noise. We set $v_\star=20$ V. To reveal the affect of measurement noise on the closed-loop system under the proposed controller, we consider two types of white noise. One white noise $w_1(t)$ has a low power and $w_2(t)$ has a high power, which are shown in Figs. \ref{Fig:noise} (a) and (c). Moreover, the curves of output voltage $v_c$ are seen in Fig. \ref{Fig:noise} (b) and (d) in the presence of $w_1(t)$ and $w_2(t)$. It is observed that the trace of output voltage has a small steady state error and slight fluctuation when $w_1(t)$ is introduced.
In addition, a higher power noise $w_2(t)$ is considered, which results in that the error and fluctuation of $v_c$ are larger. Therefore, it is concluded that the tracking performance will be deteriorated along the increase of the power of white noise, which is a common result. However, in experiment, we will adopt several techniques about software and hardware designs to reduce the affect of measurement noise on the system such that a nice tracking performance can be obtained.
}

{
{\bf Case 4}
{Here, a comparison research between the excellently robust passivity based control (RPBC) based on the differentiation at both ports and the AESC is conducted. This is also used for the control of boost converters \cite{cucuzzella2019robust}. According to this new passivity, the expression of RPBC for the buck converter with ZIP load can be given as follows
\begin{equation}\label{RPBC1}
\dot{\mu}=\frac{-K_{c}[\mu-(r^0 x_1^\star+x_2^\star)/E]-E^0 \dot{i_1}}{T_{c}}
\end{equation}
The parameters of the proposed controller are selected as $\alpha=30$, $k=3$. For the RPBC, $K_c=600000,~T_c=5000$. At 0.2 s, loads $P$, $R$, $i$ are changed from 20 W, 5 $\Omega$, 1 A to 22 W, 4 $\Omega$, 2 A, respectively.  Fig. \ref{Fig:simulation 4} shows the response curves for both controllers.  The output voltage of the closed-loop system under AESC can still quickly track the reference $v_*$ with load variations. {Due to the presence of parasitic resistance, RPBC shows poor robustness against load fluctuations.} Consequently, compared with RPBC, the AESC exhibits better convergence performance and robustness. It should be pointed out that the RPBC may be improved by adding an integral action but its stability needs to be further explored.}
}

{\bf Case 5}
{Finally, the convergence condition (\ref{nc}) is verified. The controller gains are fixed  $\alpha=10$, $k=2$. In order to display the simulation results, the convergence domain and state trajectories are represented  as projections in 3D Euclidean spaces in Fig. \ref{fig3} under nominal condition. The area surrounded by the surface is the domain satisfying the convergence condition (\ref{nc}). The curves represent state trajectories for different initial values. It can be observed that all of them can converge to the equilibrium though some initial conditions are put beyond the estimated domain.}

\begin{figure}
	\centering
	\includegraphics[scale=0.2]{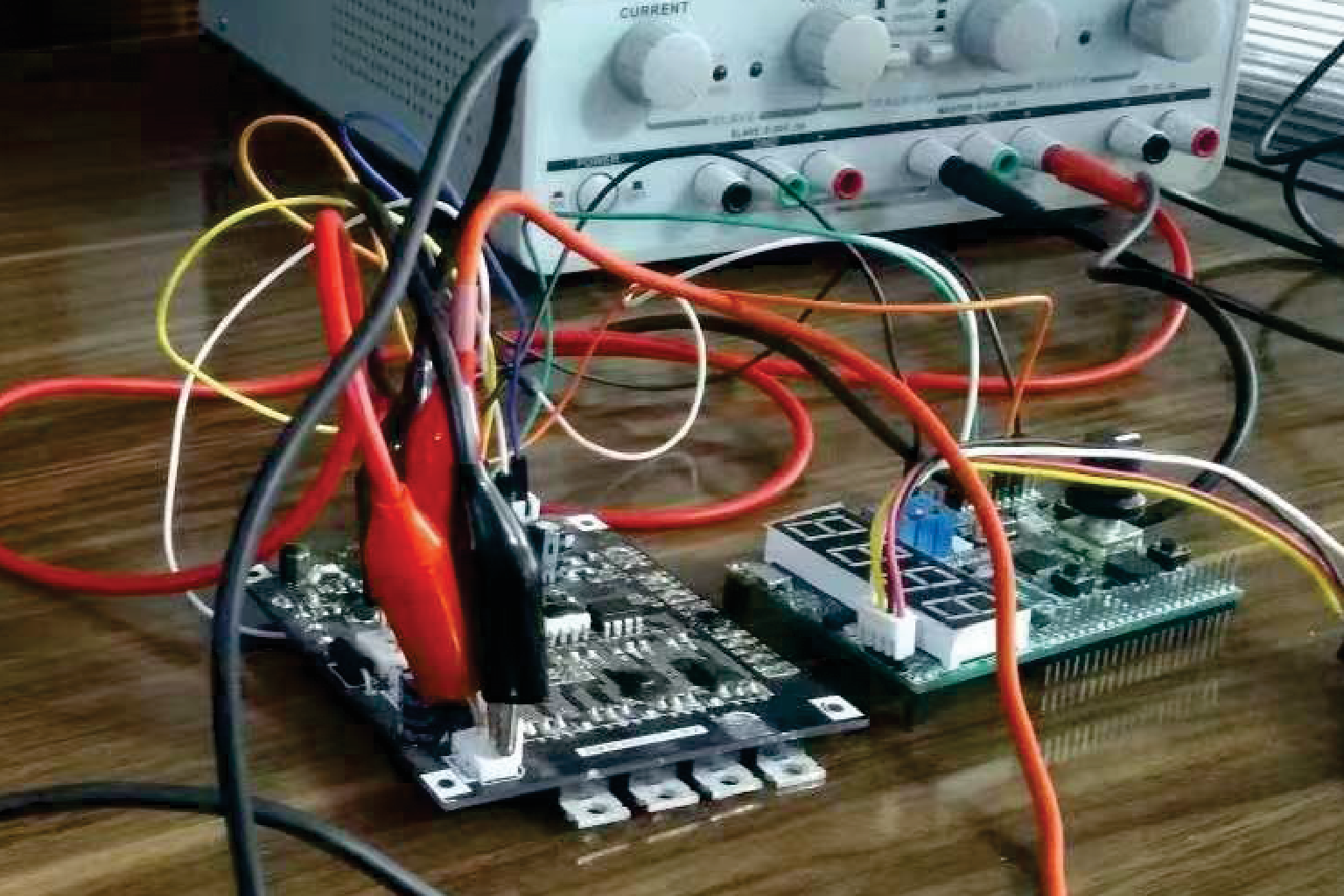}
	\caption{A photo of buck converter.}\label{setup}
\end{figure}
\section{Experimental results}\label{sec5}
In this section, the performance of designed controller will be assessed by experiment. A photo of the buck converter is shown in Fig. \ref{setup}.  {Actually, this setup contains three converters, which are buck, boost and buck-boost converters. The operation mode of the experimental setup is set by the microcontroller and different voltage levels as a buck converter.} Experimental parameters are same with them of Tab. \ref{t1}. {The power of this test is not high, that is, $30-60$ W. Due to the restriction of our experimental equipment, the operating points at higher powers are not achievable. It should be underscored that, in view of the reported theoretical analysis, it is certain that the proposed controller is also applicable to the equipments of high power and other circuit parameters provided the assumed mathematical model remains valid and that components with similar efficiencies are used. Note that low power converters may capture a certain market in smart phones, laptops, wearable devices, and medical devices.}

{
The experimental platform of buck converter is developed using the
TI MSP430F5132 MCU with the utilization of Code Composer Studio to
implement the proposed controller. The switching frequency is $100$ kHz. Signals within digital control system
are transformed to discrete-time signals described by $z$-transfer function. In terms of transfer functions MSP430F5132 enables quick data
acquisition from the various analog signals using the internal analog
to digital converter (ADC). The digital implementation of the designed
controller is based on the PWM control of the switch of buck converter. It is possible to calculate the cycle length of the C program by many ways like by measuring using the digital outputs and an oscilloscope, or using an internal timer with suitable resolution to measure code execution time. Because C itself is a low level, it is very close to the assembly, and in the assembly, read and write instructions in the register require $1$ cycle, comparison instructions require $1$ cycle, conditional branching or jump instructions require $2$ cycles, and function calls or interrupt events require $7$ cycles, approximately. Each cycle of this microcontroller is $6$ ns, and by counting the cycle of the code the execution time was less the maximum limits. Moreover, the direct memory access section of microcontroller for ADC which allows the microcontroller is adopted to perform a series of
operations without involving the CPU and the throughput of peripheral modules can be increased. In the case, the execution time of ADC will be decreased, which facilitates achieving a complicated project.
}

The CVL is constructed using a resistor with the capability of changing. The CCL is built using IRFZ44N N-channel MOSFET, and the two OPAMPS of a LM324 quad OPAMP IC, where the first OPAMP is used as voltage buffer configuration, and the second OPAMP switches the MOSFET to control the load current. The implementation of the CPL is done by adding a cascaded buck converter with resistance.
{Note that in the software design the noise in ADC module of microcontroller is mitigated by adding appropriate finite impulse response filter and also doing averaging. For the noise caused by external factors including poor PCB design and layout, electromagnetic interference or electrostatic discharge, and extreme operation environment (e.g., temperature, humidity, etc.), we can reduce the noise in the system and keeps external high frequency noise from propagating to the rest of the system by using the noise reduction techniques including the design of multi-layers PCBs compared to 2-layers ones, the uses of a capacitance between the power and ground planes and small ferrite beads placed in series between the analog power supply and digital power supply. Note that the output port of the system shown in Fig. \ref{Fig-ZIP load} is short circuited by a line in experiment.}

It is noticed that step changes in load and input voltage are considered in experiment. Then, the controller gains and the observer gains are selected as $\alpha=15,~k=2$ and $p_1(i_1)=8000 i_1$, $p_2(v_c)=100 v_c$, $p_3(i_2)=100i_2$. {Besides, we define $A_1=A_2=A_3=0$ and $M_1=M_2=M_3=1$. Moreover, in experiment, the expression of the controller $\hat \mu$ in \eqref{controller3}} is unchanged but the reference dynamics can be simplified as $\hat \mu^\star=(r^0\hat {x}_1^\star+x_2^\star-\hat d_1)/E^0, \hat {x}_1^\star=x_2^\star/R^0+P^0/x_2^\star+i^0+\hat{x}_3^\star-\hat d_2, \hat{x}_3^\star=(x_2^\star+\hat d_3)/R_2^0$ since $\lim_{t \to \infty} \dot d_1(t)=0, \lim_{t \to \infty} \dot d_2(t)=0, \lim_{t \to \infty} \dot d_3(t)=0$ are assumed.

\begin{figure}
	\centering
	\includegraphics[scale=0.35,trim=50 0 0 0,clip]{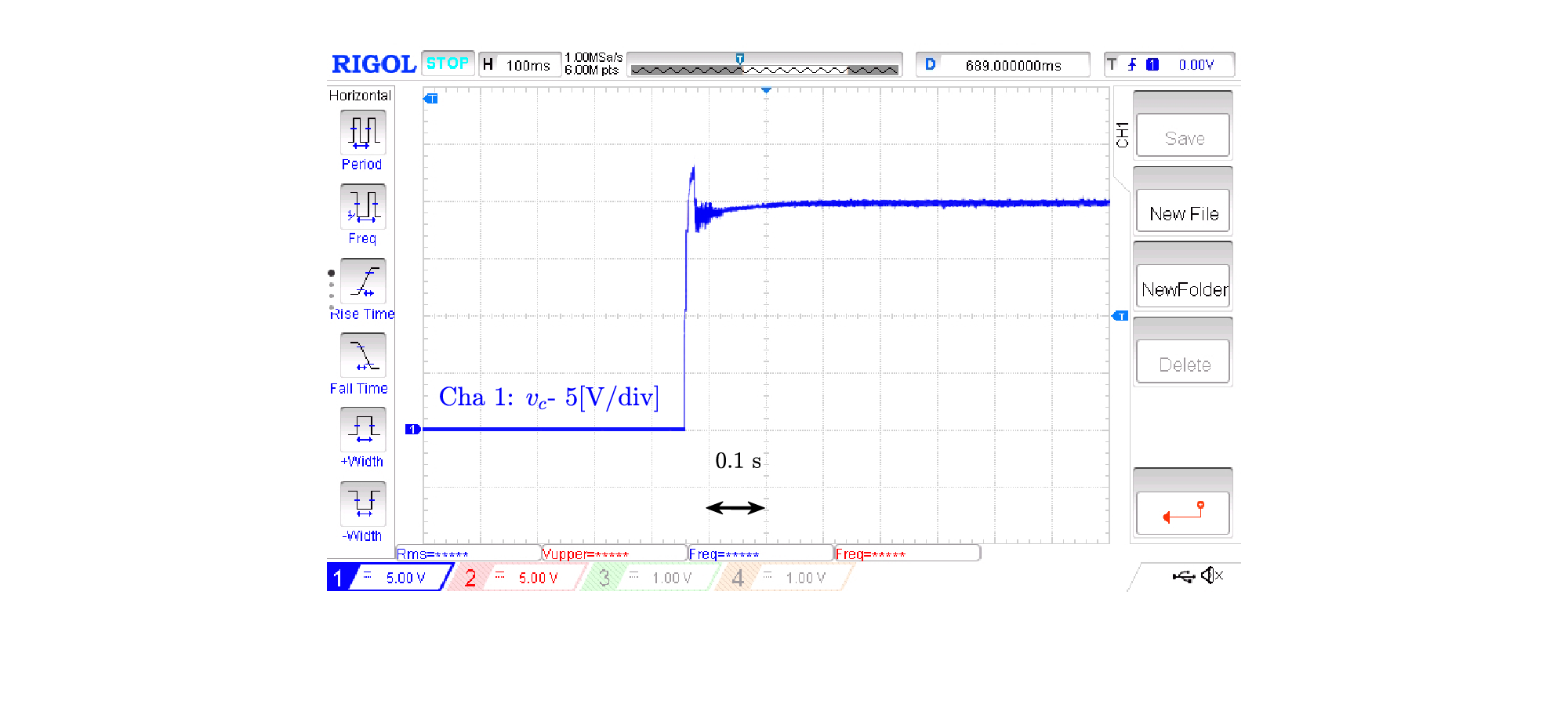}
\vspace{-0.5cm}
	\caption{The response curve of buck converter under PI controller.}\label{PI}
\end{figure}

\begin{figure}
	\centering
	\includegraphics[scale=0.35,trim=50 0 0 0,clip]{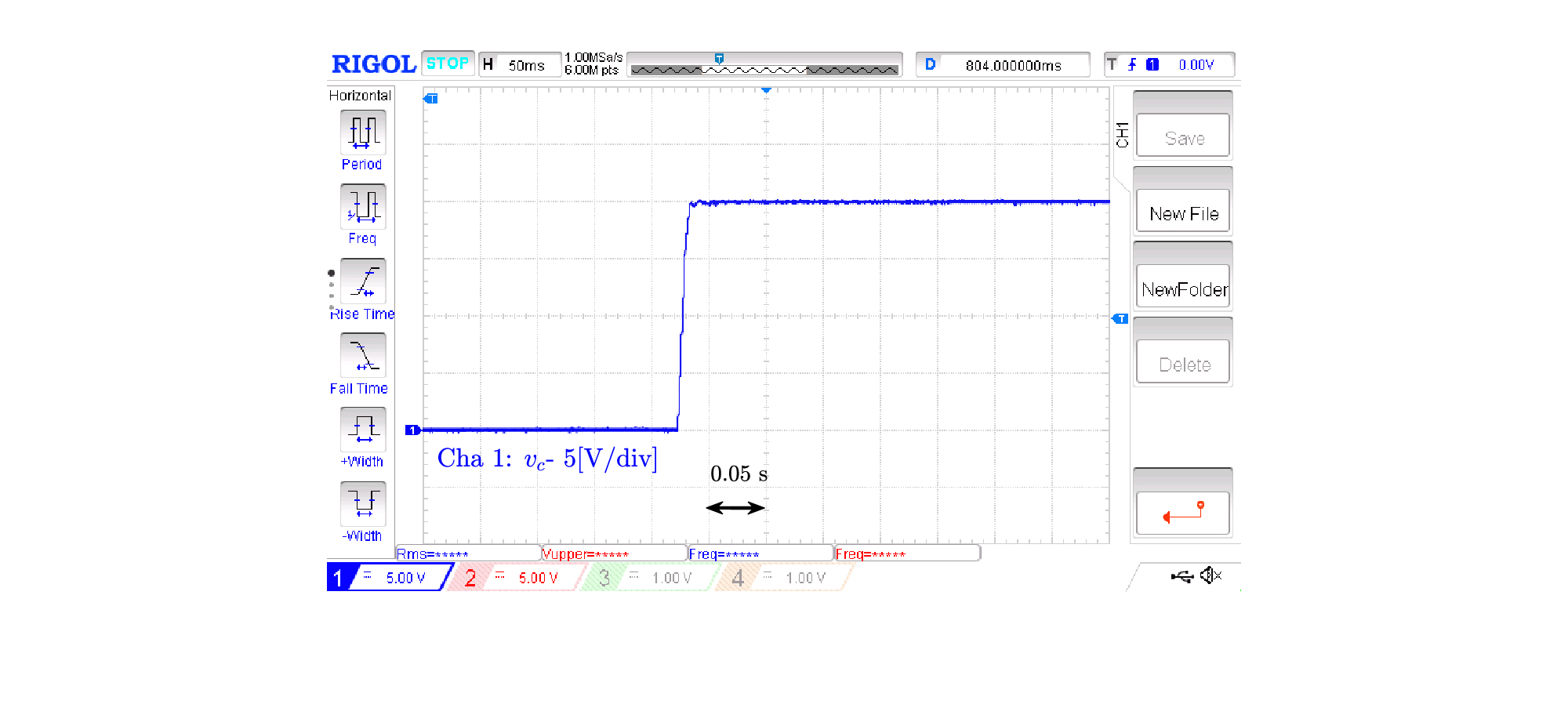}
\vspace{-0.5cm}
	\caption{The response curve of buck converter under proposed controller.}\label{PBC}
\end{figure}

{
{\bf (1) Comparison study among PI controller, RPBC and the proposed controller}. The performance comparison among PI controller, RPBC and proposed controller is conducted via experiment study. The stability analysis under PI controller is shown in \cite{he2021stabilization}.  It is observed from Figs. \ref{PI} and \ref{PBC} that the transients of proposed controller and PI controller are presented. The reference value is set as $20$ V and input voltage is fixed as $30$ V. {The comparison results are shown in Tab. \ref{t2}.} It is observed from this table that the proposed controller has a nice transient. This is similar with the simulation result, which is omitted because of the limit of the space. Note that the gains of PI control have been optimized by Matlab tool. Besides, it is seen from Figs. \ref{Fig:ex 3} and \ref{RPBC} that the proposed method can ensure output voltage $v_c$ is able to converge to the equilibrium with a small fluctuation in the presence of a step change in ZIP load. However, one sees that the RPBC obviously shows a steady state error when a step change in ZIP load occurs, which reveals a poor robustness. Moreover, its noise level seems to be higher than that of the proposed controller since the derivative of $x_1$ is used.}

\begin{table}
	\caption{The comparison between Figs. \ref{PI} and \ref{PBC}}\label{t2}
	\centering
	\begin{tabular}{lcc}
		\hline
		\hline
		Control Strategy & Overshoot & Convergence time\\
		\hline
		PI controller & 3.5 V (17.5\%) & 0.17 s\\
		AESC & 0 V  & 0.03 s\\
		\hline
		\hline
	\end{tabular}
\end{table}

\begin{figure}
	\centering
	\includegraphics[scale=0.4,trim=50 0 0 0,clip]{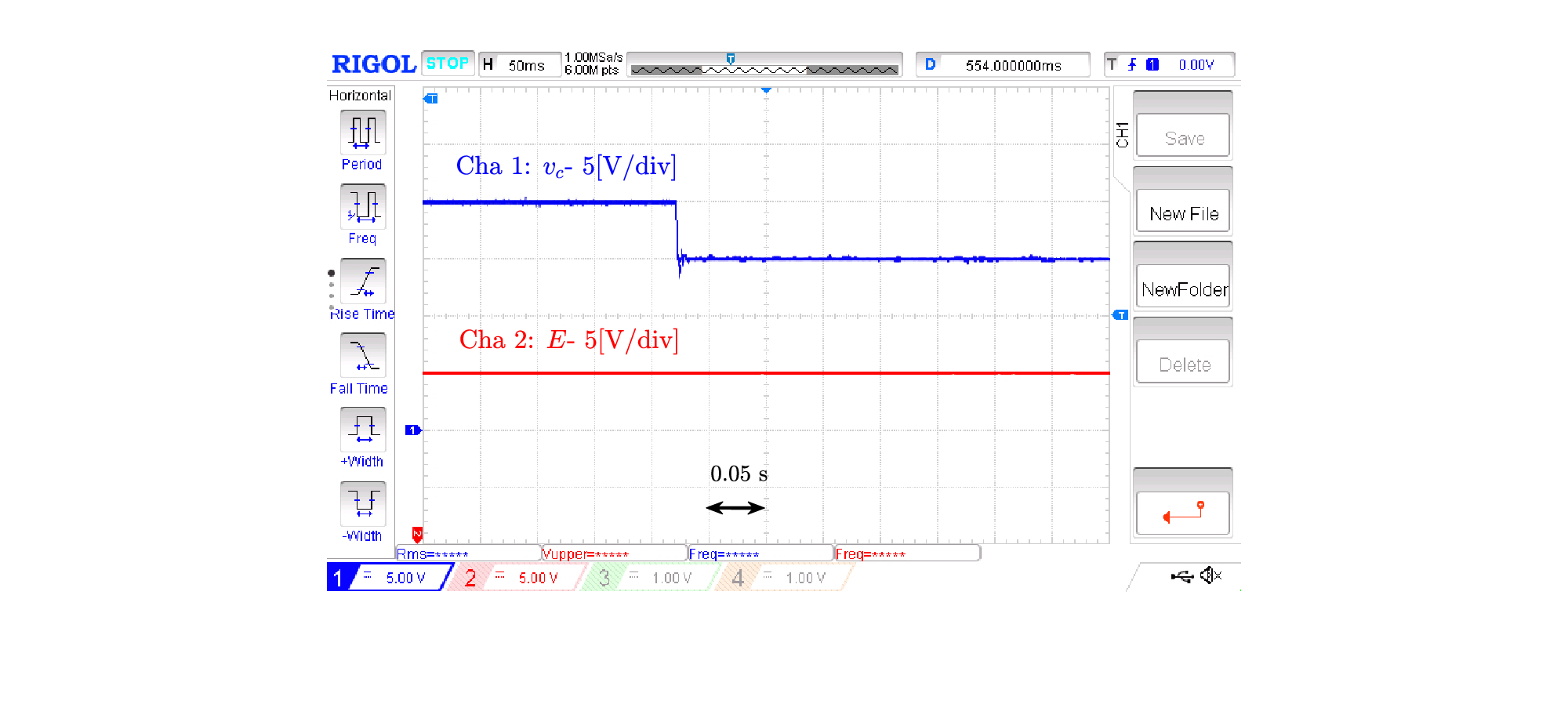}
\vspace{-0.6cm}
	\caption{The response curves of buck converter with ZIP load under a step change in reference $v_*$.}\label{Fig:ex 1}
\end{figure}

{\bf (2) Step change in reference $v_*$}.
Here, in order to test the tracking performance, the reference $v_*$ is changed from 20 V to 15 V. The response curves of Fig. \ref{Fig:ex 1} show that output voltage tracks the reference $v_*$ within 20 ms.

\begin{figure}
	\centering
	\includegraphics[scale=0.4,trim=45 0 0 0,clip]{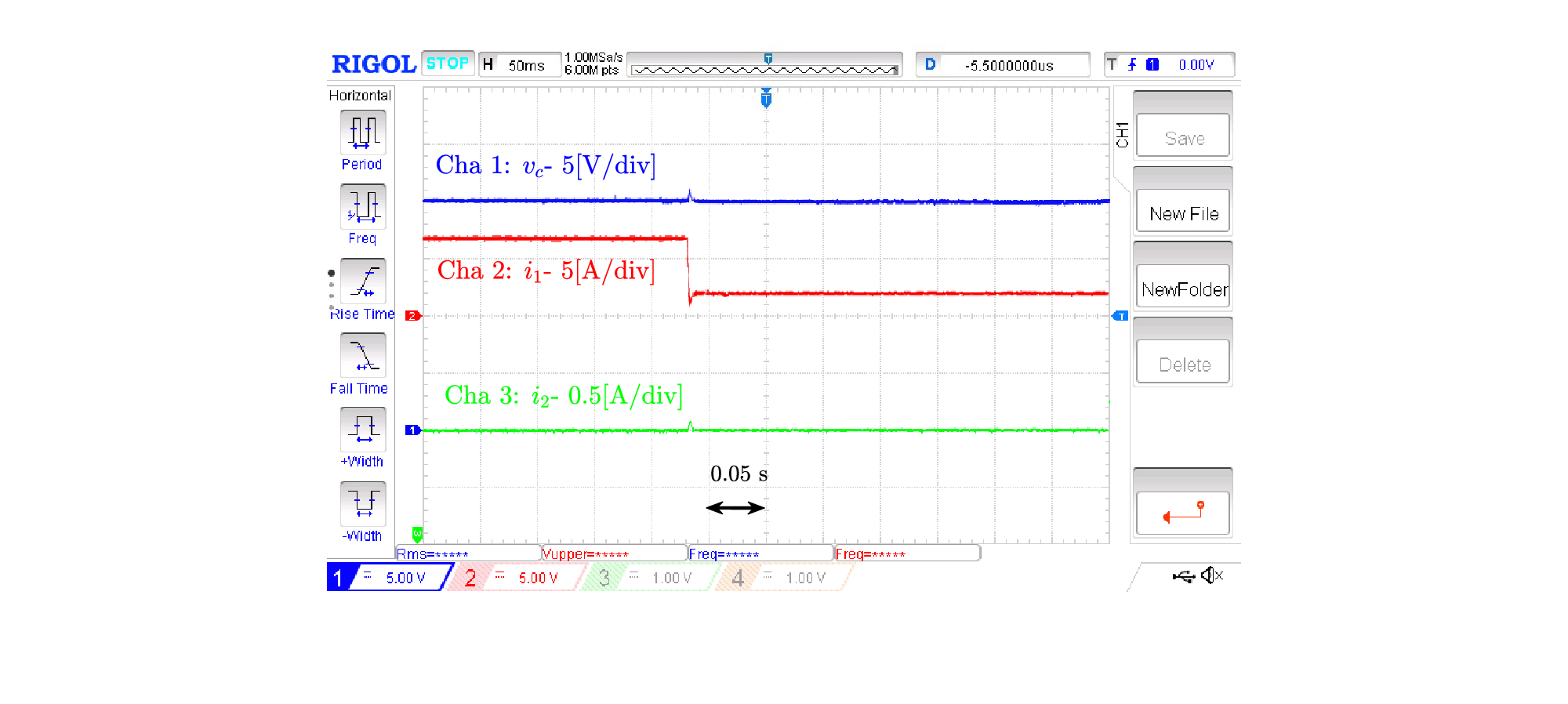}
\vspace{-0.6cm}
	\caption{The response curves of buck converter with a step change in ZIP load under AESC.}\label{Fig:ex 2}
\end{figure}

\begin{figure}
	\centering
	\includegraphics[scale=0.4,trim=45 0 0 0,clip]{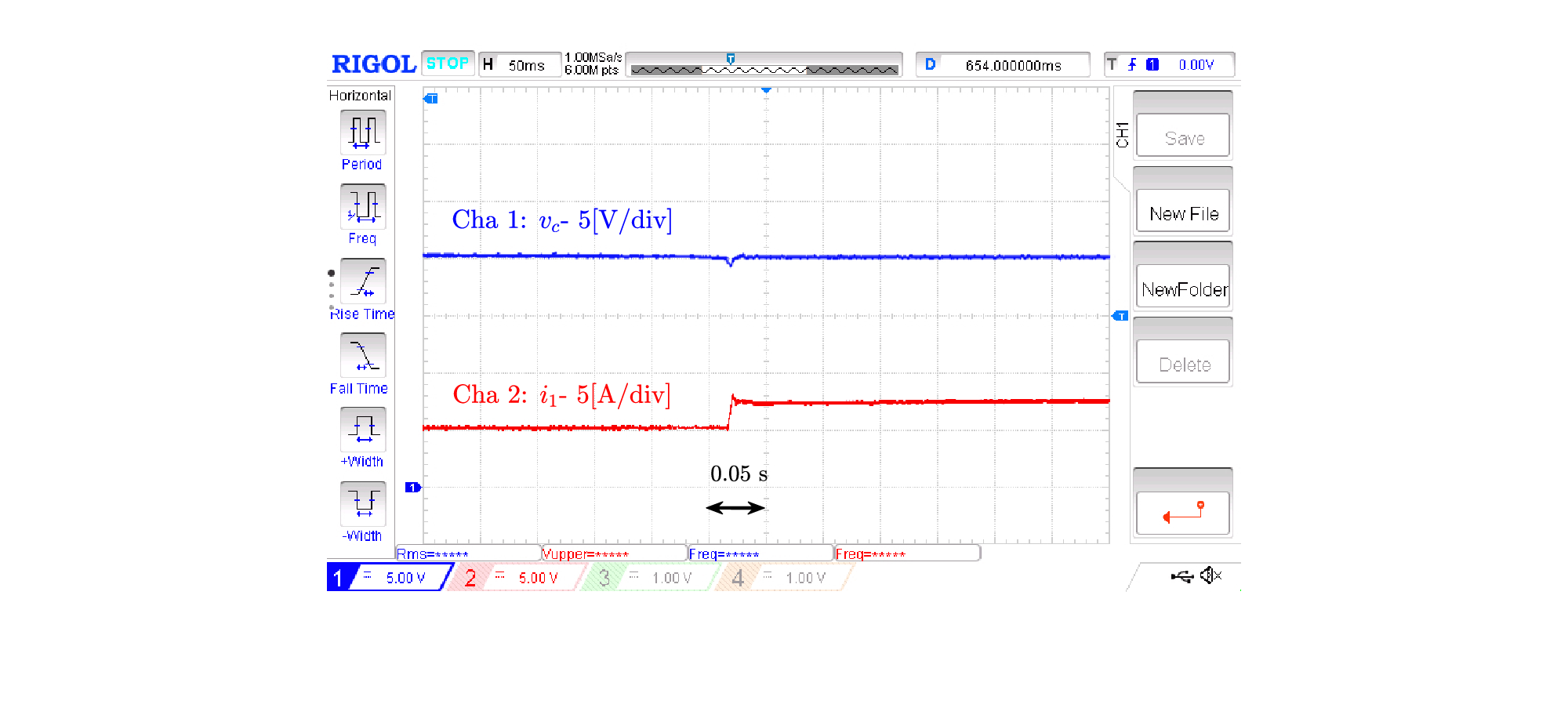}
\vspace{-0.6cm}
	\caption{The response curves of buck converter with a step change in ZIP load under AESC.}\label{Fig:ex 3}
\end{figure}

\begin{figure}
	\centering
	\includegraphics[scale=0.4,trim=45 0 0 0,clip]{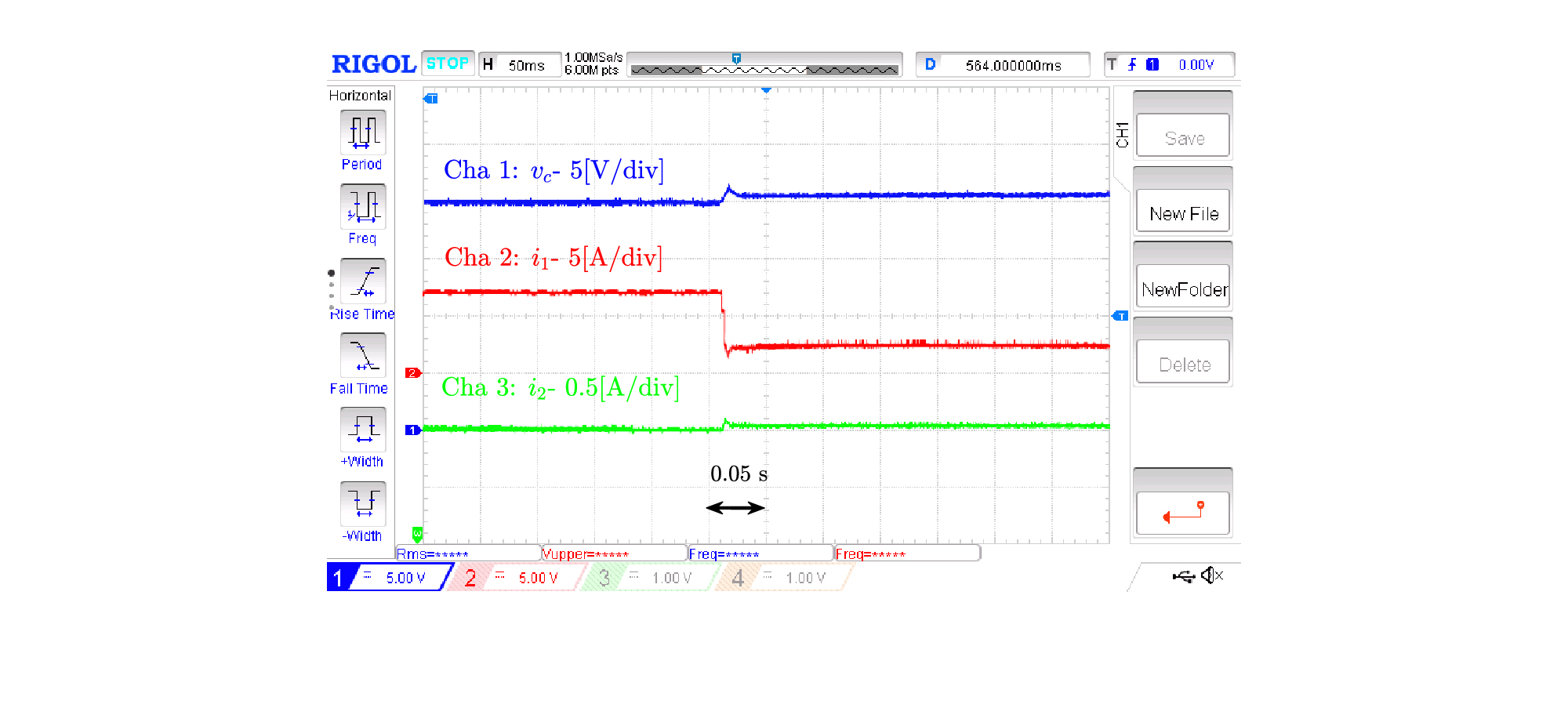}
\vspace{-0.6cm}
	\caption{The response curves of buck converter with a step change in ZIP load under RPBC \eqref{RPBC1}.}\label{RPBC}
\end{figure}

{\bf (3) Step change in ZIP load}.
Here, the effect of ZIP load disturbances on the system is investigated. First, the CVL is changed from 5 $\Omega$ to 40 $\Omega$, the CPL is reduced from 20 W to 10 W, and the CCL is decreased from 1 A to 0 A. Next, the CVL, the CPL and the CCL are changed from 10 $\Omega$, 10 W and 2 A to 5 $\Omega$, 20 W and 1 A, respectively. The response curves are revealed in Figs. \ref{Fig:ex 2} and \ref{Fig:ex 3}. It is seen from the figures that $v_c$ has only a small fluctuation with an amplitude of 1 V and recovers the reference value within 0.02 s. The influence of ZIP load disturbances on the system is effectively eliminated. Note that the response curves of the closed-loop system under the RPBC \eqref{RPBC1} are shown in Fig. \ref{RPBC} when CVL, the CPL and the CCL are changed from 5 $\Omega$, 20 W and 1 A to 40 $\Omega$, 10 W and 0 A. It is seen that load variation causes a steady state error of output voltage $v_c$, which reveals a poor robustness performance.

\begin{figure}
	\centering
	\includegraphics[scale=0.4,trim=50 0 0 0,clip]{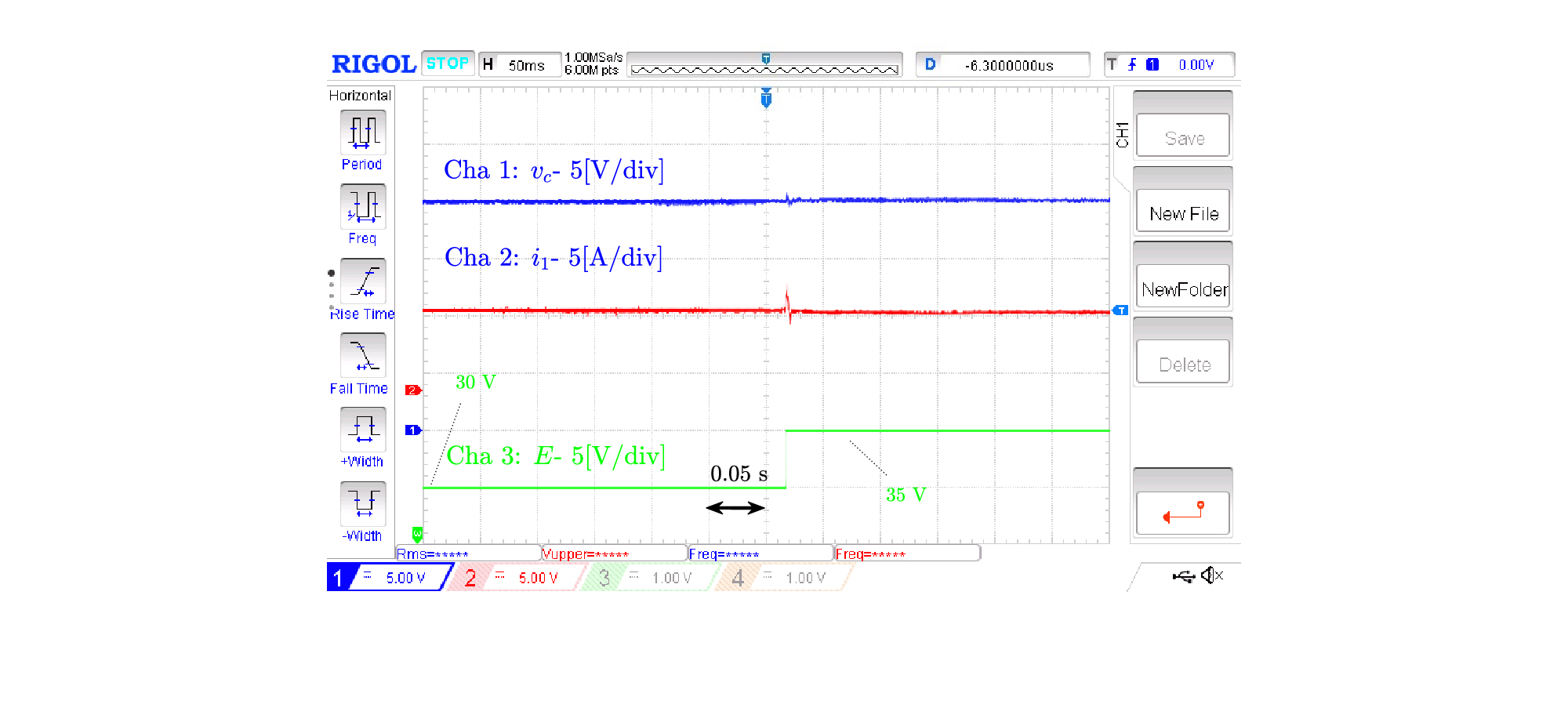}
\vspace{-0.6cm}
	\caption{The response curves of buck converter with a step change in input voltage $E$.}\label{Fig:ex 4}
\end{figure}

{\bf (4) Step change in input voltage $E$}.
The load resistance is 5 $\Omega$, the CPL is 20 W, and the CCL is 1 A. $E$ is varied from 30 V to 35 V. Fig. \ref{Fig:ex 4} displays experimental results. One sees that $v_c$ is only slightly disturbed and then stays around the reference. Hence, it is shown that the system is robust against input voltage variation.

\begin{figure}
	\centering
	\includegraphics[scale=0.55,trim=0 0 0 0,clip]{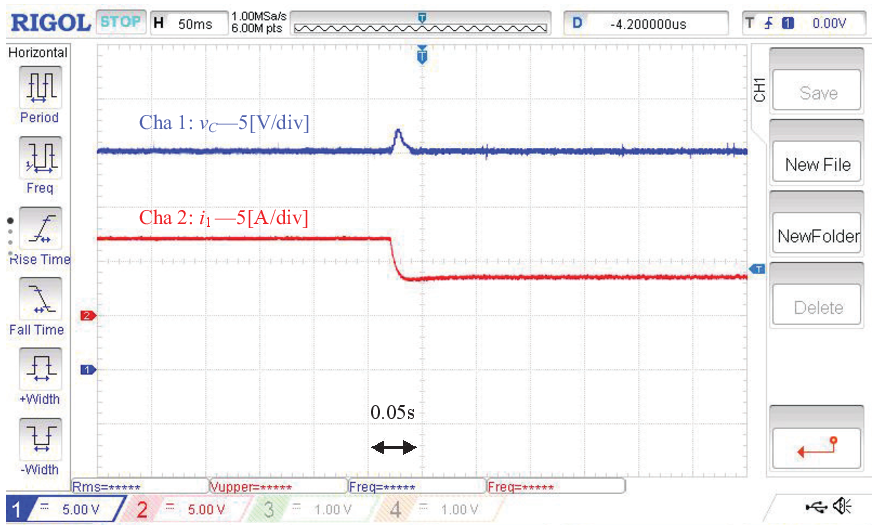}
	\caption{The response curves of buck converter with a step change in ZIP load for practical application case.}\label{Fig:ex 5}
\end{figure}

{\bf (5) Practical application consideration}. When the power line is not considered, the system is a traditional buck converter with ZIP load. Moreover, if the parasitic resistance $r$ and the perturbation of $L_1$ are neglected and step change in input voltage is considered, the proposed controller \eqref{controller} and observer \eqref{bucest2} can be simplified as $u=-\alpha \lambda k(\alpha x_1-\frac{x_c}{L_1})-\frac{\lambda x_1}{L_1}$ and ${\hat d_1}=z_1+L_1 \eta x_1, \dot {z_1}=\eta x_2-\eta \mu \hat d_1$ without needing the design of the observer for $d_2, d_3$, where $d_1=E, p_1(x_1)=\eta x_1$ and $\eta, \lambda$ are gains. This is done by slight modification and $\hat E$ is used in \eqref{transform} and $\mu^\star$ so that $\mu$ is obtained. In the case, the complicated definition and estimation of lumped disturbances can be avoided for convenience of practical application. Note that a nice robustness against load change is still obtained because of integral action. The experiment result for this consideration is shown in Fig. \ref{Fig:ex 5} with $\eta=100, \lambda=1.5, \alpha=10, k=1.5$.

\section{Conclusion}\label{sec6}
{
In this paper, we addressed the control problem of buck converter with ZIP load in the presence of matched and mismatched disturbances by designing an AESC method. A key finding lies in that an ESC is proposed with the guaranteed domain of attraction in our work. Another key finding is that the proposed observer can effectively eliminate the affect of lumped disturbances on the system so that the robustness was improved obviously. Besides, we also carried out a comparison study among the proposed controller, PI and RPBC. Simulation and experimental results were given to assess the performance of the system under three control schemes.

Future lines of research are stated as follows.
\begin{itemize}
  \item {We intend to extend other result based on energy shaping technique to deal with the control problem of boost converter feeding ZIP load. It is challenging task that a simple controller is designed for it without needing the derivatives of the states.}
  \item A generalized parameter estimation-based observer is adopted to address the current sensorless control scheme of the converter without the information of the ZIP load. This is a challenging task since the simultaneous estimate of unmeasurable state and disturbances is extremely hard. The strongly persistent excitation condition should be satisfied to ensure the convergence of the observer.
  \item {The work in \cite{ferguson2023increasing} does open a door for increasing the region of attraction for power system. We intend to use this result to report a nice result of the system with a bigger region of attraction.}
\end{itemize}
}

\bibliography{JFI}

\end{document}